\documentclass[%
reprint,
showpacs,
amsmath,amssymb,
aps,
prl
]{revtex4-1}

\usepackage{graphicx}% Include figure files
\usepackage{dcolumn}% Align table columns on decimal point
\usepackage{bm}% bold math
\begin{document}

\preprint{APS/123-QED}

\title{How ubiquitous are dragon segments in quantum transmission?}

\author{M.A.\ Novotny}
\affiliation{Dept.\ of Physics and Astronomy \\
HPC$^{\>2}$ Center for Computational Sciences \\
Mississippi State University \\
Mississippi State, MS 39762-5967 
}%

\date{\today}% It is always \today, today,

\begin{abstract}
Quantum dragon segments are nanodevices that have energy-independent total 
transmission of electrons.  
At the level of the single-band tight-binding model a nanodevice is viewed 
as a weighted undirected graph, 
with a vertex weight given by the on-site energy and the 
edge weight given by the tight-binding hopping parameter.  A quantum dragon is 
a weighted undirected graph which when connected to idealized semi-infinite 
input and output leads, has the electron transmission 
probability ${\cal T}(E)$$=$$1$.  The probability ${\cal T}(E)$ is 
obtained from the solution of the time-independent Schr{\"o}dinger equation.  
A graph must have finely tuned tight-binding parameters in 
order to have ${\cal T}(E)$$=$$1$.  
This papers addresses which weighted graphs can be tuned, by adjusting a 
small fraction of the total weights, to be a quantum dragon.  
We prove that with proper tuning any nanodevice can be a quantum dragon.  
Three prescriptions are presented to tune a weighted graph into a nanodevice. 
The implications of the prescriptions for physical nanodevices is discussed.  
\end{abstract}
%                            % Classification Scheme.
%\keywords{Quantum transmission, ballistic conduction}
%                                                %Use showkeys class option if keyword
%                             %display desired
\maketitle
%
%\tableofcontents
%
%    %    %    %    %    %    %    %    %    %    %    %    %    %    %    %    %    %
%
\section{INTRODUCTION}

Recently, the theoretical discovery of quantum dragons has been 
published \cite{QMdragon2014}.  This paper addresses the pervasiveness of 
quantum dragon nanodevices.  

All who are taught the basics of electricity learn about electrical resistance.  
Electrical resistance is the physical principle underlying  
the glow of light bulbs, 
the heating of the element in your conventional oven and of your stovetop, 
the heat produced by your laptop computers and mobile devices, 
and the energy losses in transmission lines.
A perfect electrical conductor would have zero electrical resistance, 
and consequently classically an electric current would not cause any 
heating.  

A recent discovery of a large class of nanodevices called quantum dragons 
\cite{QMdragon2014} has potential technological impacts in nano-electronics.  
A quantum dragon is a nanodevice that has zero electrical resistance, 
and therefore is a perfect conductor.  The caveat is that in the quantum regime the 
electrons propagate coherently, and for coherent propagation the 
classical concept of Ohm's law is not valid 
\cite{LAND57,DATTA1995,FerryGoodnick1997,DATTA2005,Hanson2008,NAZA2009,WONG2011}.  
A quantum dragon has zero electrical resistance in a 
four-probe measurement \cite{QMdragon2014}.  In  
a two-probe measurement, the electrical resistance of a quantum dragon 
is quantitized.  In particular, for a single open quantum channel the 
electrical resistance in a two-probe measurement 
is $G_0^{-1}$, with the quantum of conductance 
$G_0=2 e^2/h=7.748\times 10^{-5}$~S 
\cite{DATTA1995,FerryGoodnick1997,DATTA2005,Hanson2008,NAZA2009,WONG2011}.  
Here $e$ is the charge of an electron and $h$ is Planck's constant.  

The seminal work of Landauer \cite{LAND57} 
showed that from the solution of time-independent Schr{\"o}dinger equation 
\cite{Shan1994} the transmission probability as a function of energy, 
${\cal T}(E)$, was critical 
to obtain the electrical conductance via what is now called the 
Landauer formula.  
The Landauer formula is for two-probe measurements.  
The electrical conductance 
is given by a convolution of the Fermi function for the electrons 
with ${\cal T}(E)$.  
For more complicated measurements, such as a four-probe measurement, the 
Buttiker-Landauer formula is the generalization of the Landauer formula that 
must be used.  Many excellent books have been written using the Buttiker-Landauer 
formula \cite{DATTA1995,FerryGoodnick1997,DATTA2005,Hanson2008,NAZA2009,WONG2011}, 
including coherent transport through nano-structures 

\begin{figure}[tb]
\includegraphics[width=6.0cm]{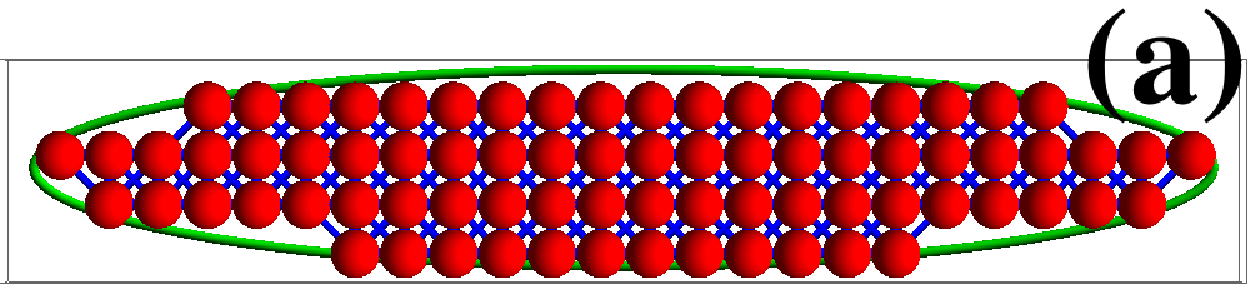}%   
\vspace{-1.5 true cm}   
\includegraphics[width=6.0cm]{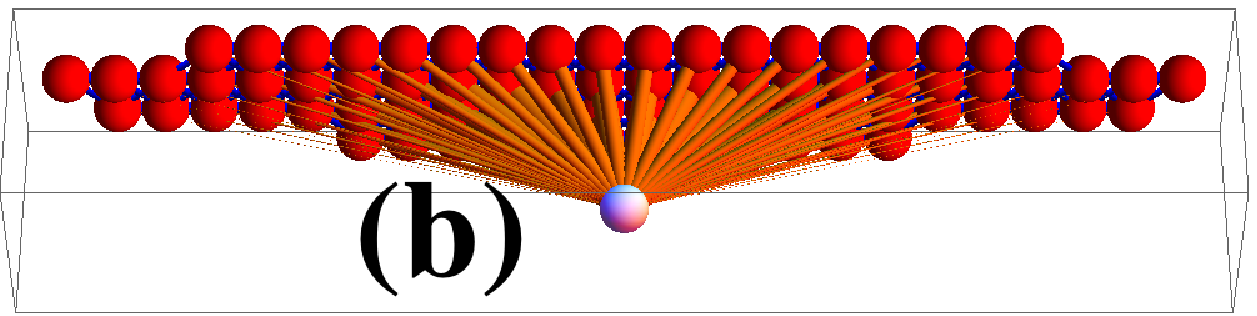}%   
\vspace{0.5 true cm}   
\includegraphics[width=6.0cm]{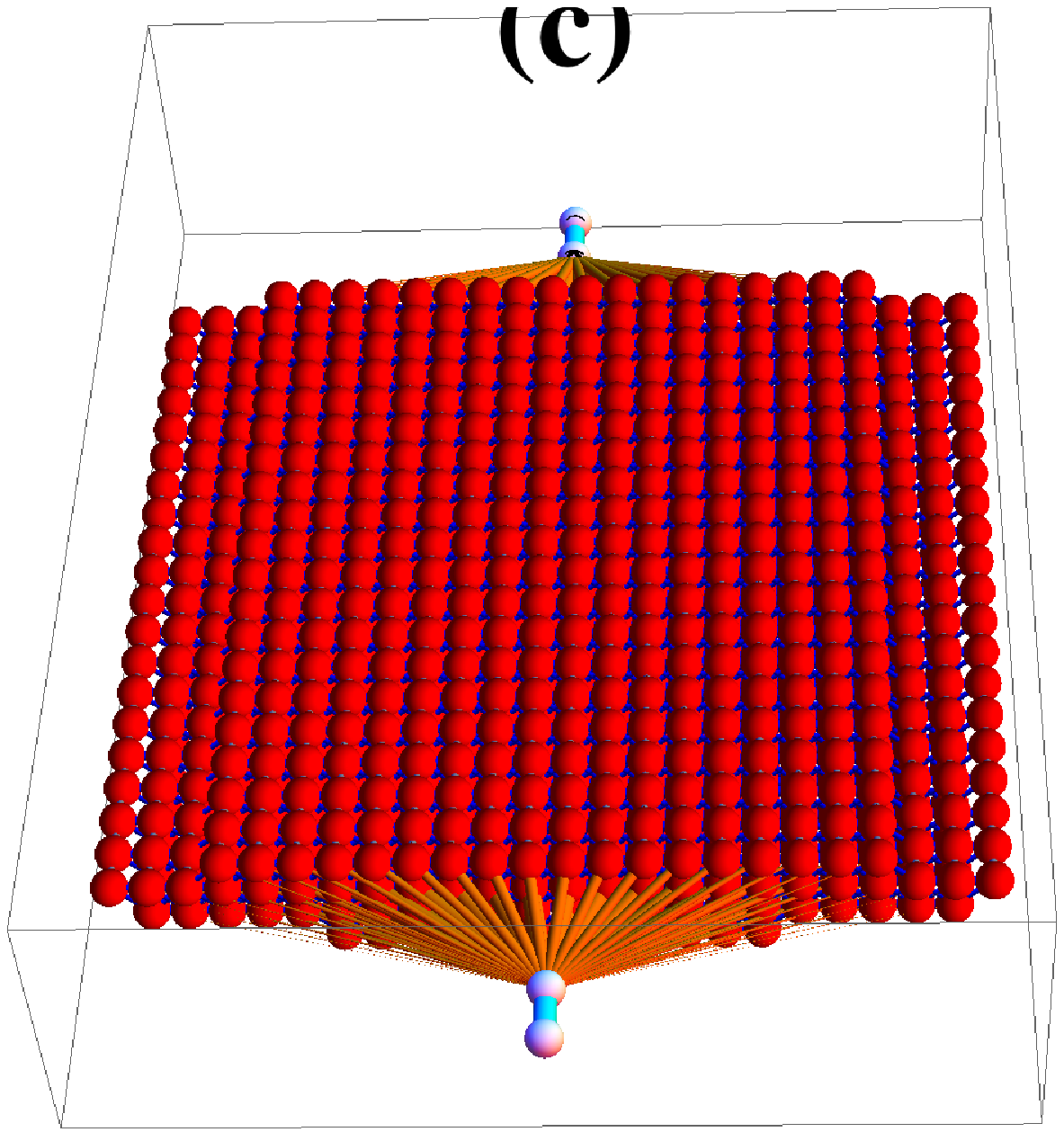}%   
\caption{\label{FigQManyDragon1} 
(Color online.)  
See the Appendix for full information on the figure construction. 
(a) An example of a slice used to make a quantum dragon, here composed of 
atoms from a square lattice within an ellipse (green curve), 
with nearest and next-nearest neighbor 
bonds (blue cylinders, which represent the intra-slice hopping terms $t_{j,j'}$).  
There are $m$$=$$76$ atoms and 236 bonds in the slice.  
(b) The hopping terms $w_j$ (orange cylinders) between a lead atom (white sphere) 
and the first slice of the nanodevice are shown.  As discussed in the text, the 
lead-device hopping terms must have particular values (denoted by the 
radius of the orange cylinders) for the nanodevice to be a quantum dragon.  
The lead atom is places above the first slice, and located at the center of mass (CM) 
of the connections $w_j$.  
(c) The complete nanodevice composed of 
$\ell$$=$$20$ identical slices, connected to the two semi-infinite 
leads (only two atoms in each lead are shown).  
}
\end{figure}

A quantum dragon is a nanodevice that when connected to idealized leads 
(as in the example of Fig.~\ref{FigQManyDragon1}) 
has full transmission of electrons for all energies that can propagate through 
the leads \cite{QMdragon2014}.  In other words, in a quantum dragon electrons for any 
propagating energy $E$ have total transmission, 
the transmission probability is ${\cal T}(E)$$=$$1$.  
Thus a quantum dragon nanodevice will behave as if the electrons 
underwent ballistic propagation.  However, ballistic propagation requires 
no scattering of incident electrons, while there may be strong scattering of 
electrons in a quantum dragon.  The best known experimental 
example of a quantum dragon is a single-walled carbon nanotube in the armchair 
configuration 
\cite{Mintmire1992,HAMADA1992,Frank1998,Dresselhaus2004,Tans1997,Berger2002}.  
For a quantum dragon ${\cal T}(E)$$=$$1$, and in a two-probe 
measurement the Landauer formula thus gives the single-channel 
electrical conductance $G_0$.  

At the most basic level, the single-band tight-binding model for electrical 
transmission can be viewed as a weighted undirected graph \cite{BUSA1965,CHART2012} 
connected to idealized 
semi-infinite leads.  The tight-binding parameters include the 
on-site energy associated with a the vertex weight (atom weight or node weight) 
of the graph.  Vertex $j$ will have a weight $\epsilon_j$.  
The tight-binding hopping parameters are the weights given to the 
edges (bonds) of the graph.  
%The semi-infinite leads considered here will 
%be taken to have all vertices with on-site energy zero (setting the zero of 
%the energy scale) and bond weights of strength unity (setting the energy unit).  
We consider graphs composed of $\ell$$\ge$$1$ slices, with only electron 
hopping allowed within a slice or between atoms in nearest-neighbor slices.  
We use $t_{j,j'}$ as the strength of the intra-slice hopping parameters 
between vertices $j$ and $j'$.  Similarly, we use  
$s_{j,j'}$ as the strength of the inter-slice hopping parameters between 
vertices.  Similarly, $w_{j}$ or $u_{j}$ are the strengths of the weights of 
hopping between the last (first) vertex of the input (output) lead and each 
vertex in the first (last) slice.  
The statements and proofs in this paper are in regard to weighted 
undirected graphs.  
The discussion section will bring the weighted undirected graph study 
back to examination of physical nanodevices.  

In this paper, we show the weighted undirected graph associated with 
{\it any\/} nanodevice can be a quantum dragon.  
The question to be addressed is how many parameters within the tight-binding model 
must be tuned in order to obtain ${\cal T}(E)$$=$$1$.  
Three prescriptions are presented to make the weighted graphs into 
a quantum dragon.  
In the first two prescriptions, we consider only homogeneous weighted graphs made of 
$\ell$ identical slices of $m$ atoms, 
with only the simplest inter-slice connections.  
In the third prescription the constraint of homogeneity of slices is relaxed.  
The three prescriptions to make a weighted undirected 
graph a quantum dragon by tuning certain tight-binding parameters are listed below.  
In all cases we only consider connected graphs for the nanodevice, 
although some of these prescriptions may be modified to deal 
with graphs that are not connected.  
\begin{enumerate}
\item If all $\epsilon_i$ and intra-slice hopping $t_{j,j'}$ are arbitrarily fixed, 
and only the simplest inter-slice hopping parameters $s_{j.j'}$ are included, 
a weighted graph composed of identical slices can be a quantum dragon 
by tuning $2m$$+$$2$ parameters:
\begin{itemize}
\item The hopping parameters between the lead and 
nanodevice must each be a unique value; 
\item A uniform electrical potential shift of all on-site energies $\epsilon_j$; 
\item All inter-slice hopping parameters $s_{j,j'}$ must be scaled (tuned) by the 
same value; 
\item The output lead connections to the last ($\ell^{\rm th}$) slice 
are identical to the input lead connections to the first slice.  
\end{itemize}
\item If all input lead to nanodevice hopping parameters, $w_j$, 
are in arbitrarily fixed ratios 
and non-zero, only the simplest inter-slice hopping terms are included, 
and all $\epsilon_j$ and intra-slice hopping $t_{j,j'}$ are fixed, 
requires tuning $2m$$+$$2$ parameters:
\begin{itemize}
\item Each $\epsilon_j$ must be shifted (tuned) by a unique value of the 
electrical potential at site $j$ by $V_j$;
\item The strength of all lead-device hopping parameters must be normalized 
(tuned) in a unique fashion; 
\item All inter-slice hopping parameters $s_{j,j'}$ must be scaled (tuned) by the 
same value; 
\item The output lead connections, $u_j$, to the last ($\ell^{\rm th}$) slice 
are identical to the input lead connections, $w_j$, to the first slice.  
\end{itemize}
\item If all $\epsilon_j$ and intra-slice hopping $t_{j,j'}$ are arbitrarily fixed in 
each slice, but the slices of the weighted graph are not homogeneous: 
\begin{itemize}
\item The hopping parameters between the leads and 
the nanodevice must each be a unique (tuned) value, which may be 
different for the connections to the input lead ($w_{j}$) and 
output lead ($u_{j}$); 
\item A uniform electrical potential shift (tuning) 
of all on-site energies $\epsilon_j$ in each slice, 
but the tuning required may be different for each slice; 
\item A unique (tuned) value for each of the inter-slice hopping terms 
$s_{j,j'}$, which may be different between each pair of atoms on adjacent slices.  
\end{itemize}
\end{enumerate}

The vast majority of calculations use the Green's function 
method to calculate the electrical conductance 
\cite{DATTA1995,FerryGoodnick1997,DATTA2005,Hanson2008,NAZA2009,WONG2011}.  
The Green's function implicitly solves the Schr{\"o}dinger equation, and 
through the Landauer equation \cite{LAND57} gives ${\cal T}(E)$ and the 
electrical conductivity.  
In this paper, we use a different standard technique, called the 
matrix method \cite{DCA2000}, which solves the 
Schr{\"o}dinger equation using an particular {\it ansatz\/}.  
From the matrix method solution of the Schr{\"o}dinger equation, 
one can calculate ${\cal T}(E)$, and hence the electrical conductivity from 
the Landauer equation \cite{LAND57}.  
There are several papers that use the matrix method to calculate ${\cal T}(E)$
\cite {QMdragon2014,DCA2000,Lin03,ISLAM2008,CUANSING2009,BOET2011,Varghese2012}.  
The physical quantity, ${\cal T}(E)$, is independent of which calculation method 
is used since they both satisfy the time-independent Schr{\"o}dinger equation 
for the semi-infinite leads connected to the nanodevice.  

%         %         %         %         %         %         %         %         %
%         %         %         %         %         %         %         %         %

%
%    %    %    %    %    %    %    %    %    %    %    %    %    %    %    %    %    %
%
\section{TIGHT-BINDING MODEL}

A simple example of a weighted undirected graph associated with a 
physically realizable nanodevice that may be made into a quantum dragon 
is shown in Fig.~\ref{FigQManyDragon1}.  A simple-cubic (SC) lattice structure 
is presented in Fig.~\ref{FigQManyDragon1}.  
The metal polonium (Po) has a simple cubic 
lattice structure, and hence Fig.~\ref{FigQManyDragon1} can be viewed as a 
nano-crystal of Po.  
Fig.~\ref{FigQManyDragon1}(a) shows a single slice of the nanodevice, made up 
of all atomic sites in a square lattice that fit into a given ellipse.  
The center of the ellipse is randomly given within a square lattice unit cell, 
leading to a fixed arbitrary non-isotropic atomic arrangement.  
The first slice of the nanodevice is connected to the input lead ($w_{j}$), 
tuned as shown 
in Fig.~\ref{FigQManyDragon1}(b).  
The last slice of the nanodevice is connected to the output lead in an 
identical fashion, namely $u_j$$=$$w_j$ for all $j$.  
An example of a twenty slice ($\ell$$=$$20$) nanodevice 
connected to the leads is shown in Fig.~\ref{FigQManyDragon1}(c), 
one which may be tuned to be a quantum dragon.

We assume that the on-site energy of the semi-infinite leads is zero, 
thereby setting our zero of energy.  We take the hopping parameters 
between the lead atoms as $-1$, thereby setting our energy scale as 
the strength of the lead-lead hopping term.  
We take the lattice spacing within the leads to be unity, 
setting the unit for length.  
With these units, only electrons with energies 
$-2\le E\le 2$ propagate in the leads \cite{QMdragon2014,DATTA1995,DCA2000}.  

For the tight-binding model the hopping parameters are obtained from the 
kinetic energy terms of the Schr{\"o}dinger equation \cite{DATTA1995}, and therefore 
must be non-positive.  That is the reason for the negative sign in 
the hopping parameter $-1$ within the leads.  
We will take all hopping parameters ($t_{j,j'}$, $s_{j,j'}$, $w_j$, and $u_j$) 
to be positive, {\it i.e.\/} the hopping strengths are positive.  
The required negative signs are put in explicitly, so $t_{j,j'}$$\ge$$0$ and 
$s_{j,j'}$$\ge$$0$.  

%
%    %    %    %    %    %    %    %    %    %    %    %    %    %    %    %    %    %
%
\section{TRANSMISSION VIA THE MATRIX METHOD}

In order the calculate the transmission of an incoming electron 
as a function of energy, ${\cal T}(E)$, the time-independent Schr{\"o}dinger 
equation needs to be solved.  The matrix method for solving the 
tight-binding model is used here \cite{DCA2000}.  
This method has been used and published a sufficient number of times 
that the derivation of the equations \cite{QMdragon2014} is not required.  
Therefore, only the relevant equations are given in order to set the 
notation \cite{QMdragon2014} and give the reader the basics of the matrix method.  

The nanodevice is composed of $\ell$ slices.  
For the first two quantum dragon prescriptions, the slices are 
identical, each with $m$ atoms.  
The equation to solve for the Schr{\"o}dinger equation 
of the device (with $\ell m$ vertices) and the semi-infinite leads is infinite.  
After using an {\it ansatz\/} for the leads, the final matrix equation 
has a linear dimension $\ell m$$+$$2$ \cite{QMdragon2014,DCA2000}.  
We assume that only atoms in nearest-neighbor slices interact.  
Fig.~\ref{FigQManyDragon1} shows an example with $m$$=$$76$ and 
$\ell$$=$$20$.  
For $\ell$$=$$4$ slices, the matrix equation to analyze has the form 
\begin{equation}
\label{Eq:EqN4}
{\bf N}_4 = 
\left(
\begin{array}{cccccc}
\xi(E) & {\vec w}^\dagger & {\vec 0}^\dagger & 
                            {\vec 0}^\dagger & {\vec 0}^\dagger & 0 \\
{\vec w} & {\bf F} & {\bf B} & {\bf 0} & {\bf 0} & {\vec 0} \\ 
{\vec 0} & {\bf B}^\dagger & {\bf F} & {\bf B} & {\bf 0} & {\vec 0} \\ 
{\vec 0} & {\bf 0} & {\bf B}^\dagger & {\bf F} & {\bf B} & {\vec 0} \\ 
{\vec 0} & {\bf 0} & {\bf 0} & {\bf B}^\dagger & {\bf F} & {\vec u} \\ 
0 & {\vec 0}^\dagger & {\vec 0}^\dagger & {\vec 0}^\dagger & 
                            {\vec u}^\dagger & \xi(E) \\
\end{array}
\right)
\> .
\end{equation}
The definition 
$\xi(E)=\left(-E-i\sqrt{4-E^2}\right)/2$ has been made.  
The quantity $\xi(E)$ can be viewed as the result of coupling the 
finite graph associated with the nanodevice to the semi-infinite leads.  
The zero matrices, ${\bf 0}$, are $m$$\times$$m$, and the zero vectors 
${\vec 0}$ have $m$ elements.   
The inter-slice hopping matrix ${\bf B}$ is $m$$\times$$m$ and has elements of the 
inter-slice hopping parameters.  For the simplest inter-slice connections, 
as shown in Fig.~\ref{FigQManyDragon1}(c), 
the inter-slice hopping matrix is given by ${\bf B}=-s_0{\bf I}$ where 
${\bf I}$ is the $m$$\times$$m$ identity matrix.  
The vector ${\vec w}$ has as its elements the hopping parameters 
between the last atom of the input lead and the atoms in the first slice.  
The vector ${\vec u}$ has as its elements the hopping parameters 
between the first atom of the output lead and the atoms in the 
last ($\ell^{\rm th}$) slice.  
The $m$$\times$$m$ matrix ${\bf F}={\bf A}-E{\bf I}$, with $E$ the 
energy of the incoming electron.  
The $m$$\times$$m$ matrix ${\bf A}$ has as its $j^{\rm th}$ diagonal element the 
on-site energy $\epsilon_j$ of the $j^{\rm th}$ atom (vertex).  
The $j,j'$ off-diagonal element of ${\bf A}$ is the intra-slice hopping term 
$-t_{j,j'}$.  
The matrix ${\bf A}$ is symmetric, since the Schr{\"o}dinger equation involves 
a Hamiltonian, and we here restrict ourselves to real values for the 
on-site energies and all the hopping parameters.  
In Fig.~\ref{FigQManyDragon1}(a) the intra-slice hopping terms are shown as 
cylinders, with the radius of the cylinder representing the strength of the 
intra-slice hopping term $t_{j,j'}$.  

The matrix equation to solve for the transmission is given by 
(again written for $\ell$$=$$4$ slices) 
\begin{equation}
\label{Eq:N4equation}
{\bf N}_4
\left(
\begin{array}{c}
1 + r \\
{\vec \psi}_1 \\
{\vec \psi}_2 \\
{\vec \psi}_3 \\
{\vec \psi}_4 \\
t_T \\
\end{array}
\right)
= 
\left(
\begin{array}{c}
\Xi \\
{\vec 0} \\
{\vec 0} \\
{\vec 0} \\
{\vec 0} \\
0 \\
\end{array}
\right)
\> ,
\end{equation}
with the definition 
$\Xi(E) = -2 i \> {\cal I}{\it m}\left(\xi(E)\right)$ \cite{QMdragon2014,DCA2000}.  
The wavefunction of slice $j$ is given by ${\vec\psi}_j$.  
For a given energy of the incoming electron, 
the inverse of the matrix of ${\bf N}_4$ in Eq.~(\ref{Eq:N4equation}) is 
calculated.  This enables one to obtain the wavefunctions ${\vec\psi}_j$, 
as well as $r$ and $t_T$.  Note that these all depend on the energy $E$ of the 
incoming electron.  
The quantity $t_T$ should not be confused with the 
hopping parameters in ${\bf N}_\ell$; 
unfortunately both are conventionally denoted by a lower case $t$, so the 
subscript $T$ denotes that $t_T$ is for the transmission, not a hopping parameter.  
The transmission probability of the electron is given by 
\begin{equation}
{\cal T}(E)=\left|t_T(E)\right|^2
\> .  
\end{equation}
The reflection probability for an electron is given by 
${\cal R}$$=$$\left|r\right|^2$.  Every electron is either reflected or 
transmitted, so ${\cal R}$$+$${\cal T}$$=$$1$.  

%
%    %    %    %    %    %    %    %    %    %    %    %    %    %    %    %    %    %
%
\section{PROJECTION MAPPING TECHNIQUE}

In order to more easily calculate $t_T(E)$ in 
Eq.~(\ref{Eq:N4equation}) we introduce a transformation 
matrix ${\hat{\bf X}}$, written for $\ell$$=$$3$, 
\begin{equation}
\label{Eq:Xhat}
{\hat{\bf X}}_\ell = \left(
\begin{array}{ccccc}
1 & {\vec 0}^\dagger & {\vec 0}^\dagger & {\vec 0}^\dagger & 0 \\
{\vec 0} & {\bf X} & {\bf 0} & {\bf 0} & {\vec 0} \\
{\vec 0} & {\bf 0} & {\bf X} & {\bf 0} & {\vec 0} \\
{\vec 0} & {\bf 0} & {\bf 0} & {\bf X} & {\vec 0} \\
0 & {\vec 0}^\dagger & {\vec 0}^\dagger & {\vec 0}^\dagger & 1 \\
\end{array}
\right)
\end{equation}
with the $m$$\times$$m$ matrices ${\bf X}$ unitary 
({\it i.e.\/} ${\bf X}^\dagger{\bf X}$$=$${\bf I}$). 
Note that (written for $\ell$$=$$3$) 
\begin{equation}
\label{Eq:hatXrightVector}
{\hat{\bf X}}_3 
\left(
\begin{array}{c}
\Xi \\
{\vec 0} \\
{\vec 0} \\
{\vec 0} \\
0 \\
\end{array}
\right)
=
\left(
\begin{array}{c}
\Xi \\
{\vec 0} \\
{\vec 0} \\
{\vec 0} \\
0 \\
\end{array}
\right)
\>.
\end{equation}

Form the matrix 
${\bf M}_\ell = {\hat{\bf X}}_\ell{\bf N}_\ell {\hat{\bf X}}_\ell^\dagger$.  
In the standard fashion, by matrix multiplication 
the $m$$\times$$m$ diagonal matrix blocks in ${\bf M}_\ell$ are given by 
${\bf X}{\bf F}{\bf X}^\dagger$.  
The non-zero off-diagonal blocks of ${\bf M}_\ell$ are given by 
${\bf X}{\bf B}{\bf X}^\dagger$ and ${\bf X}{\vec w}$, or by the complex congugates.  
For example, for $\ell$$=$$3$
\begin{equation}
{\bf M}_3 \> = \> 
\left(
\begin{array}{ccccc}
\xi(E) & {\vec w}^\dagger{\bf X}^\dagger & {\vec 0}^\dagger & {\vec 0}^\dagger & 0 \\
{\bf X}{\vec w} & {\bf X}{\bf F}{\bf X}^\dagger & {\bf X}{\bf B}{\bf X}^\dagger 
                & {\bf 0} & {\vec 0} \\
{\vec 0} & {\bf X}{\bf B}^\dagger{\bf X}^\dagger & {\bf X}{\bf F}{\bf X}^\dagger 
                & {\bf X}{\bf B}{\bf X}^\dagger & {\vec 0} \\
{\vec 0} & {\bf 0} & {\bf X}{\bf B}^\dagger{\bf X}^\dagger 
                & {\bf X}{\bf F}{\bf X}^\dagger & {\bf X}{\vec u} \\
0 & {\vec 0}^\dagger & {\vec 0}^\dagger & {\vec u}^\dagger{\bf X}^\dagger & \xi(E) \\
\end{array}
\right)
\>.
\end{equation}
Multiplying Eq.~(\ref{Eq:N4equation}) on the left by ${\hat{\bf X}}_\ell$, 
inserting ${\hat{\bf X}}^\dagger_\ell {\hat{\bf X}}_\ell$$=$${\bf I}$ 
between ${\bf N}_\ell$ and the vector in Eq.~(\ref{Eq:N4equation}) 
that contains the wavefunctions, 
and using Eq.~(\ref{Eq:hatXrightVector}) gives (written for $\ell$$=$$4$) 
\begin{equation}
\label{Eq:N4M4equation}
{\hat{\bf X}}_4{\bf N}_4
{\hat{\bf X}}_4^\dagger \> 
{\hat{\bf X}}_4
\left(
\begin{array}{c}
1 + r \\
{\vec \psi}_1 \\
{\vec \psi}_2 \\
{\vec \psi}_3 \\
{\vec \psi}_4 \\
t_T \\
\end{array}
\right)
=
{\bf M}_4 
\left(
\begin{array}{c}
1 + r \\
{\bf X}{\vec \psi}_1 \\
{\bf X}{\vec \psi}_2 \\
{\bf X}{\vec \psi}_3 \\
{\bf X}{\vec \psi}_4 \\
t_T \\
\end{array}
\right)
%= 
%{\hat{\bf X}}_4
%\left(
%\begin{array}{c}
%\Xi \\
%{\vec 0} \\
%{\vec 0} \\
%{\vec 0} \\
%{\vec 0} \\
%0 \\
%\end{array}
%\right)
= 
\left(
\begin{array}{c}
\Xi \\
{\vec 0} \\
{\vec 0} \\
{\vec 0} \\
{\vec 0} \\
0 \\
\end{array}
\right)
\> .
\end{equation}

We have complete freedom to choose the unitary transformation matrices ${\bf X}$.  
Assume we can find a ${\bf X}$ that satisfies the 
four mapping equations \cite{QMdragon2014} 
\begin{equation}
\label{Eq:Map1}
{\bf X}{\bf A}{\bf X}^\dagger = 
\left(
\begin{array}{cc}
{\tilde\epsilon} & {\vec 0}^\dagger \\
{\vec 0} & {\tilde {\bf A}} \\
\end{array}
\right)
\> ,
\end{equation}
\begin{equation}
\label{Eq:Map2}
{\bf X}{\bf B}{\bf X}^\dagger = 
\left(
\begin{array}{cc}
-{\tilde s}_b & {\vec 0}^\dagger \\
{\vec 0} & {\tilde {\bf B}} \\
\end{array}
\right)
\> ,
\end{equation}
\begin{equation}
\label{Eq:Map3}
{\bf X}{\vec w} = 
\left(
\begin{array}{c}
-{\tilde s}_w \\
{\vec 0} \\
\end{array}
\right)
\>,
\end{equation}
and
\begin{equation}
\label{Eq:Map4}
{\bf X}{\vec u} = 
\left(
\begin{array}{c}
-{\tilde s}_u \\
{\vec 0} \\
\end{array}
\right)
\> .
\end{equation}
The $(m$$-$$1)$$\times$$(m$$-$$1)$ matrices ${\tilde{\bf A}}$ and ${\tilde{\bf B}}$ 
are not important, since they will not be connected by any path to either the 
input or output leads \cite{QMdragon2014}.  

Introduce the $(\ell$$+$$2)$$\times$$(\ell$$+$$2)$ matrix 
${\widetilde{\bf M}}_\ell$, written for $\ell$$=$$4$, as 
\begin{equation}
\label{Eq:widetildeM4}
{\widetilde{\bf M}}_4 = 
\left(
\begin{array}{cccccc}
\xi(E) & -{\tilde s}_w & 0 & 0 & 0 & 0 \\
-{\tilde s}_w & {\tilde\kappa} & -{\tilde s}_b & 0 & 0 & 0 \\
0 & -{\tilde s}_b & {\tilde\kappa} & -{\tilde s}_b & 0 & 0 \\
0 & 0 & -{\tilde s}_b & {\tilde\kappa} & -{\tilde s}_b & 0 \\
0 & 0 & 0 & -{\tilde s}_b & {\tilde\kappa} & -{\tilde s}_u \\
0 & 0 & 0 & 0 & -{\tilde s}_u & \xi(E) \\
\end{array}
\right)
\end{equation}
with ${\tilde\kappa}={\tilde\epsilon}-E$.  
This is the matrix formed from 
only the $\ell$ transformed-sites in Eq.~(\ref{Eq:N4M4equation}) 
that are connected to the leads, after the mapping equations 
Eq.~(\ref{Eq:Map1}) through Eq.~(\ref{Eq:Map4}) are used.  

The probability of transmission of the electron of energy $E$, 
${\cal T}$$=$$\left|t_T\right|^2$ 
are calculated from $t_T$ found from either the equation 
(written for $\ell$$=$$4$, with the displayed vectors of length $4m$$+$$2$) 
\begin{equation}
\label{Eq:MappedMbig}
\left(
\begin{array}{c}
1+r \\
{\bf X} {\vec\psi}_1 \\
{\bf X} {\vec\psi}_2 \\
{\bf X} {\vec\psi}_3 \\
{\bf X} {\vec\psi}_4 \\
t_T
\end{array}
\right)
= 
{\bf M}_4^{-1} 
\left(
\begin{array}{c}
-2 i \> {\cal I}{\it m}(\xi) \\
{\vec 0} \\
{\vec 0} \\
{\vec 0} \\
{\vec 0} \\
0 \\
\end{array}
\right)
\end{equation}
or from the equation
(written for $\ell$$=$$4$, with the displayed vectors of length $6$$=$$4$$+$$2$) 
\begin{equation}
\label{Eq:MappedMsmall}
\left(
\begin{array}{c}
1+r \\
{\tilde\phi}_1 \\
{\tilde\phi}_2 \\
{\tilde\phi}_3 \\
{\tilde\phi}_4 \\
t_T
\end{array}
\right)
= 
{\widetilde{\bf M}}_4^{-1} 
\left(
\begin{array}{c}
-2 i \> {\cal I}{\it m}(\xi) \\
0 \\
0 \\
0 \\
0 \\
0 \\
\end{array}
\right)
\>,
\end{equation}
where the ${\tilde\phi}_j$ are the first element of the vector 
${\bf X}{\vec\psi}_j$.  
The mapping equations of Eq.~(\ref{Eq:Map1}) through (\ref{Eq:Map4})
thus reduce substantially the size of the matrix that one must find the 
inverse of in order to calculate $t_T$.  
The matrix 
${\bf M}_\ell$ is $(\ell m$$+$$2)$$\times$$(\ell m$$+$$2)$ while the 
matrix ${\widetilde{\bf M}}_\ell$ is 
$(\ell$$+$$2)$$\times$$(\ell$$+$$2)$.  
Note that provided the mapping equations 
[Eq.~(\ref{Eq:Map1}) through (\ref{Eq:Map4})]
hold, no approximation is made by 
going from Eq.~(\ref{Eq:MappedMbig}) to Eq.~(\ref{Eq:MappedMsmall}), 
as they both give identical transmissions 
${\cal T}(E)$$=$$\left| t_T\right|^2$.  

%
%    %    %    %    %    %    %    %    %    %    %    %    %    %    %    %    %    %
%
\section{QUANTUM DRAGONS FROM MAPPING}

Recall that our lead sites have on-site energies set to zero and a hopping 
of strength unity.  
The nanodevice will be a quantum dragon, {\it i.e.\/} will have 
${\cal T}(E)$$=$$\left| t_T\right|^2$$=$$1$, if 
${\tilde\epsilon}$$=$$0$ and 
${\tilde s}_w$$=$${\tilde s}_u$$=$${\tilde s}_b$$=$$1$.  
The reason for the complete transmission of electrons of all energies is 
that these are the values that the matrix in 
Eq.~(\ref{Eq:widetildeM4}) would have if one used the matrix method 
to calculate the transmission through a homogeneous infinite wire, 
but selected a string of $\ell$ lead sites to be a nanodevice.  Since 
the wire is homogeneous, there is no scattering by the nanodevice, and 
the electrons for any energy that propagate through the lead would be 
completely transmitted.  Note, however, that the original slices 
(as in Fig.~\ref{FigQManyDragon1}) may be 
very inhomogeneous, which may lead to strong scattering of the electrons.  

There is complete freedom in terms of the transformation matrices 
${\bf X}$ that are used.  The only requirement to go from 
Eq.~(\ref{Eq:MappedMbig}) to Eq.~(\ref{Eq:MappedMsmall}) is that the 
four mapping equations,
Eq.~(\ref{Eq:Map1}) through (\ref{Eq:Map4}), are satisfied.  
For the first two quantum dragon prescriptions, 
only nanodevices with the simplest inter-slice coupling between identical slices 
(as in Fig.~\ref{FigQManyDragon1}(c)) will be analyzed.  
This means the inter-slice hopping matrix is ${\bf B}$$=$$-s_b{\bf I}$.  
Hence Eq.~(\ref{Eq:Map2}) is satisfied since ${\bf X}$ is unitary.  
Thus one has for these two prescriptions 
${\tilde s}_b$$=$$s_b$.  In order to have the nanodevice have the possibility 
of being a quantum dragon we tune the parameter $s_b$$=$$1$.  
Therefore, for such a quantum dragon nanodevice we need only to find a 
transformation matrix ${\bf X}$ that satisfies 
Eqs.~(\ref{Eq:Map1}), (\ref{Eq:Map3}) and (\ref{Eq:Map4}), and that have 
${\tilde\epsilon}$$=$$0$ and ${\tilde s}_w$$=$${\tilde s}_u$$=$$1$.  

The third quantum dragon prescription will require the four mapping 
equations, Eqs.~(\ref{Eq:Map1}) through (\ref{Eq:Map4}) to be 
satisfied, and to have 
for each slice $k$ the mapped on-site energy ${\tilde\epsilon}_k$$=$$0$, 
between each nearest-neighbor pair of slices ${\tilde s}_{k,k+1}$$=$$1$, 
between the input lead and the first slice ${\tilde s}_{w}$$=$$1$, 
and between the output lead and the last ($\ell^{\rm th}$) 
slice ${\tilde s}_{u}$$=$$1$.  

%
%    %    %    %    %    %    %    %    %    %    %    %    %    %    %    %    %    %
%
\section{QUANTUM DRAGONS: PRESCRIPTION~1}

We consider a weighted undirected graph made from 
$\ell$ identical slices each with $m$ atoms.  
As in Fig.~\ref{FigQManyDragon1} the weighted graph may be associated with the 
tight-binding model on a physical nanodevice.  
We assume the intra-slice matrix elements ${\bf A}$ are all fixed to arbitrary values 
(both the on-site energies $\epsilon_j$ and intra-slice hopping terms $t_{j,j'}$).  
We assume that the slice atoms are strongly connected, in the sense of 
graph theory \cite{BUSA1965,CHART2012}.  
In other words, within a slice every atom can be visited starting from 
any other atom by a series of hops using only the non-zero intra-slice hopping 
terms $t_{j,j'}$.  
We assume that only the simplest inter-slice connections are present, 
so ${\bf B}=-s_0{\bf I}$.  
We assume we are free to tune all the lead-slice 
connection strengths in ${\vec w}$ and ${\vec u}$, 
to add a constant electric potential $V$ to the on-site energy of every slice 
(which is the same for every atom in the nanodevice), 
and to tune the strength of the inter-slice hopping strengths $s_0$.  

Let $\epsilon_{\rm max}$ be the maximum of zero or the 
largest positive diagonal element of ${\bf A}$, {\it i.e.\/} 
\begin{equation}
\label{Eq:Rx1first}
\epsilon_{\rm max}={\rm max}\{0,\epsilon_1,\epsilon_2,\cdots,\epsilon_m\}
\end{equation}
with the $\epsilon_j$ the on-site energies, which are the diagonal elements 
of the matrix ${\bf A}$.  
Introduce the matrix
\begin{equation}
{\hat{\bf A}} = \epsilon_{\rm max} {\bf I} -{\bf A}
.  
\end{equation}
Then the matrix ${\hat{\bf A}}$ is non-negative 
\cite{MarcusMinc64,BermanPlemmons1979}, which is 
written as ${\hat{\bf A}}\ge 0$.  In other words, every element of 
${\hat{\bf A}}$ is positive or zero.  
Let ${\vec w}_1$ be the normalized eigenvector of ${\hat{\bf A}}$ associated with 
the largest eigenvalue ${\hat\lambda}_1$ of ${\hat{\bf A}}$, {\it i.e.\/} 
\begin{equation}
{\hat{\bf A}}{\vec w}_1={\hat\lambda}_1 {\vec w}_1
\>.
\end{equation}
Since ${\hat{\bf A}}$ has an associated strongly connected graph and is non-negative, 
by a well-known extension of the Perron-Frobenius theorem 
\cite{MarcusMinc64,BermanPlemmons1979}
the vector ${\vec w}_1$ is unique and can be written with all 
non-negative values (${\vec w}_1$$\ge$$0$).  
Then ${\vec w}_1$ is also an eigenvector of ${\bf A}$ with eigenvalue 
$\lambda_1$, {\it i.e.\/} 
\begin{equation}
\label{Eq:Pres1Aeigen}
{\bf A}{\vec w}_1=\lambda_1 {\vec w}_1
\>. 
\end{equation}
Remember that the eigenvalues are functions of all the elements of ${\bf A}$, 
including the largest diagonal element of ${\bf A}$, {\it i.e.\/} 
\begin{equation}
\lambda_1 = \lambda_1\left(\epsilon_{\rm max}\right) 
= \epsilon_{\rm max}-{\hat\lambda}_1\left(\epsilon_{\rm max}\right)
\>.
\end{equation}
Note that ${\bf A}$ and 
${\hat{\bf A}}$ are symmetric, 
so that they can be associated with an undirected graph, 
as opposed to being associated with a directed graph as would be required for general 
non-negative matrices.  
Since ${\bf A}$ is symmetric, ${\vec w}_1$ is both a right and left 
eigenvector of ${\bf A}$.  

We tune the lead-device hopping terms, which must all be non-positive, to be 
\begin{equation}
{\vec w} = -{\hat s}_w {\vec w}_1
\quad {\rm and}\quad
{\vec u} = -{\hat s}_w {\vec w}_1
\end{equation} 
with some overall strength ${\hat s}_w$ that we will determine below.  
The strengths are the same for the connections to the input and to the output leads.
We have had to form the matrix ${\hat{\bf A}}$ in order to use the 
Perron-Frobenius theorem to show that one can obtain vectors ${\vec w}$ and 
${\vec u}$ with all non-positive values.  
This is the physical constraint imposed by the hopping being the negative of 
the kinetic energy portion of the time-independent Schr{\"o}dinger equation.  

Choose the transformation matrix to be (written for $m$$=$$5$) 
\begin{equation}
\label{Eq:Xw5}
{\bf X}^\dagger = 
\left(
\begin{array}{ccccc}
\> {\vec w}_1 \>\> & 
{\vec w}_2 \>\> & 
{\vec w}_3 \>\> & 
{\vec w}_4 \>\> & 
{\vec w}_5 \>\> \\
\end{array}
\right)
\end{equation}
with orthonormal vectors ${\vec w}_j^\dagger {\vec w}_{j'}$$=$$\delta_{j,j'}$.  
The vector ${\vec w}_1$ is an eigenvector of ${\bf A}$ from 
Eq.~(\ref{Eq:Pres1Aeigen}).  The 
other vectors ${\vec w}_j$ for $j$$=$$2, 3, \cdots, m$ 
need not be eigenvectors of ${\bf A}$, only orthonormal to each other and 
to ${\vec w}_1$.  
With this choice of ${\bf X}$ and ${\vec w}$, 
the mapping equation Eq.~(\ref{Eq:Map3}) is 
satisfied with ${\tilde s}_w$$=$${\hat s}_w$.  
Similarly, the mapping equation Eq.~(\ref{Eq:Map4}) is 
satisfied with ${\tilde s}_u$$=$${\hat s}_w$.  
The connections to the input and output leads are the same since all 
slices are identical.  

We tune one parameter by applying a constant electrical potential $-V_{\rm shift}$ 
to every atom in the nanodevice, so the matrix ${\bf A}$ is shifted 
to the matrix ${\bf A}$$-$$V_{\rm shift}{\bf I}$.  
Note the electric potential shift is the same for all atoms in the nanodevice.  
Then 
\begin{equation}
\left(
{\bf A} - V_{\rm shift}{\bf I}
\right) {\vec w}_1 =
\left(
\lambda_1 - V_{\rm shift}
\right) {\vec w}_1 
\> .
\end{equation}
With our choice of ${\bf X}$, the mapping equation Eq.~(\ref{Eq:Map1}) is 
satisfied with 
\begin{equation}
{\tilde\epsilon} = 
\lambda_1(\epsilon_{\rm max}) - V_{\rm shift}
\>.
\end{equation}

All four mapping equations have thus been satisfied.  
For a quantum dragon using this prescription, we need only choose 
\begin{equation}
\label{Eq:Rx1last}
\begin{array}{lcl}
{\vec w} & \> = \> & - {\vec w}_1 \\
{\vec u} & \> = \> & - {\vec w}_1 \\
& {\rm and} & \\
V_{\rm shift} & \> = \> & \lambda_1 \\
\end{array}
\end{equation}
so that ${\tilde s}_w$$=$${\tilde s}_u$$=$$1$ and ${\tilde\epsilon}$$=$$0$.  
We have tuned the $m$ lead-slice interactions in 
${\vec w}=-{\vec w}_1$, 
the $m$ lead-slice interactions in ${\vec u}=-{\vec w}_1$, 
the constant shift electrical potential 
$V_{\rm shift}$, and the inter-slice hopping strength $s_b$$=$$1$.  
In all, we have tuned $2m$$+$$2$ tight-binding parameters.  
With the simplest inter-slice connections ${\bf B}$$=$$-s_0{\bf I}$, 
the total number of possible parameters in the tight-binding model 
is equal to 
\begin{equation}
\label{Eq:pres1Nterms}
\begin{array}{lcl}
N_{{\rm Total}\>{\rm parameters}} 
& \>=\> &
\frac{m\left(m-1\right)}{2} + 3m +1 \\
& \> = \> &
\frac{m^2+5m+2}{2}
\>. \\
\end{array}
\end{equation}
The number of intra-slice hopping terms $t_{j,j'}$ is $m(m-1)/2$, 
the number of different on-site energies $\epsilon_j$ is $m$, 
the number of lead-slice hopping terms is $m$ for the input lead and 
$m$ for the output lead, and there is the 
inter-slice hopping strength $s_0$, giving the result in 
Eq.~(\ref{Eq:pres1Nterms}).  

\begin{figure}[tb]
\includegraphics[width=6.5cm]{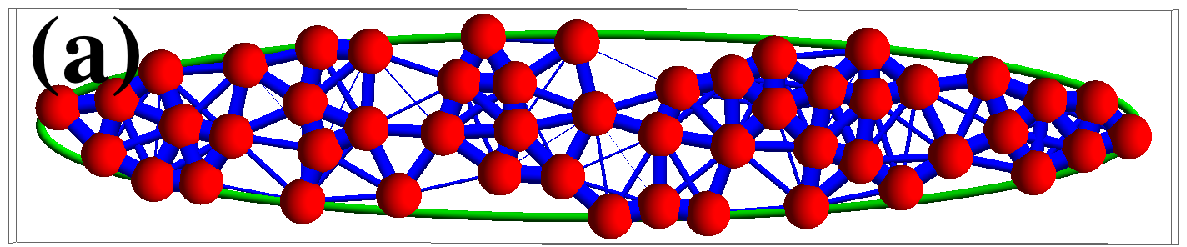}%   
\hskip 1.3 true in   
\includegraphics[width=6.5cm]{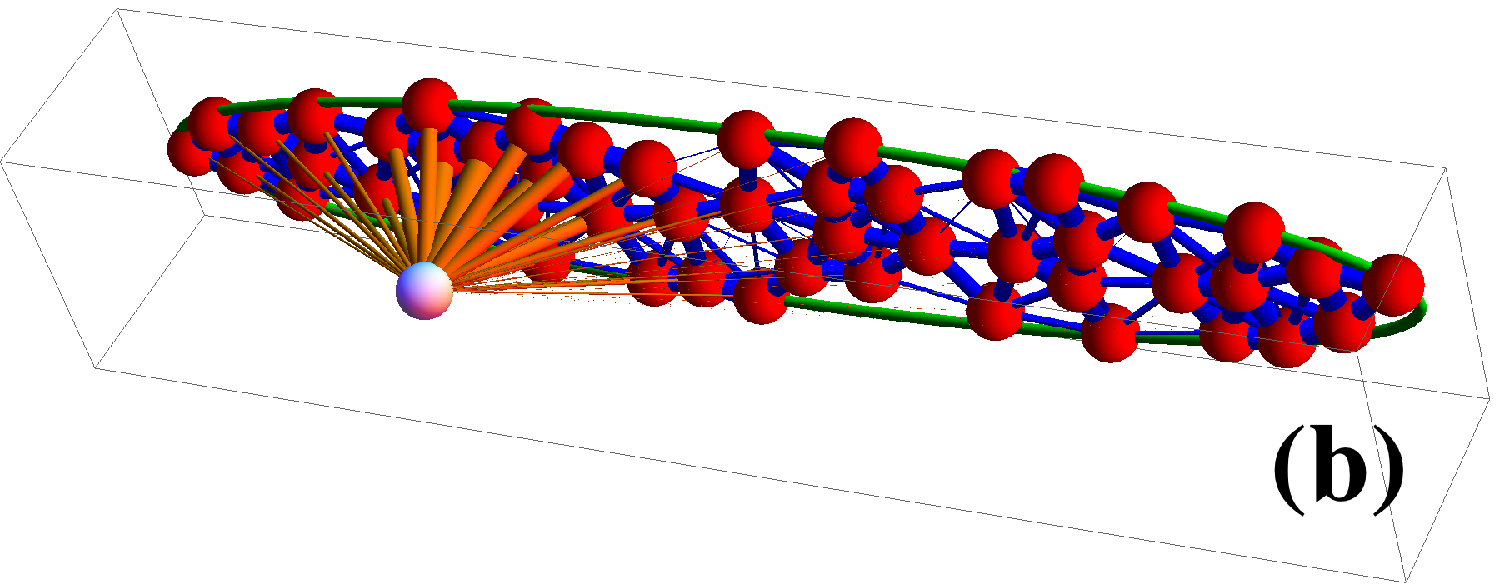}%   
\hskip 1.3 true in   
\includegraphics[trim=300 0 300 0,clip,width=6.5cm]{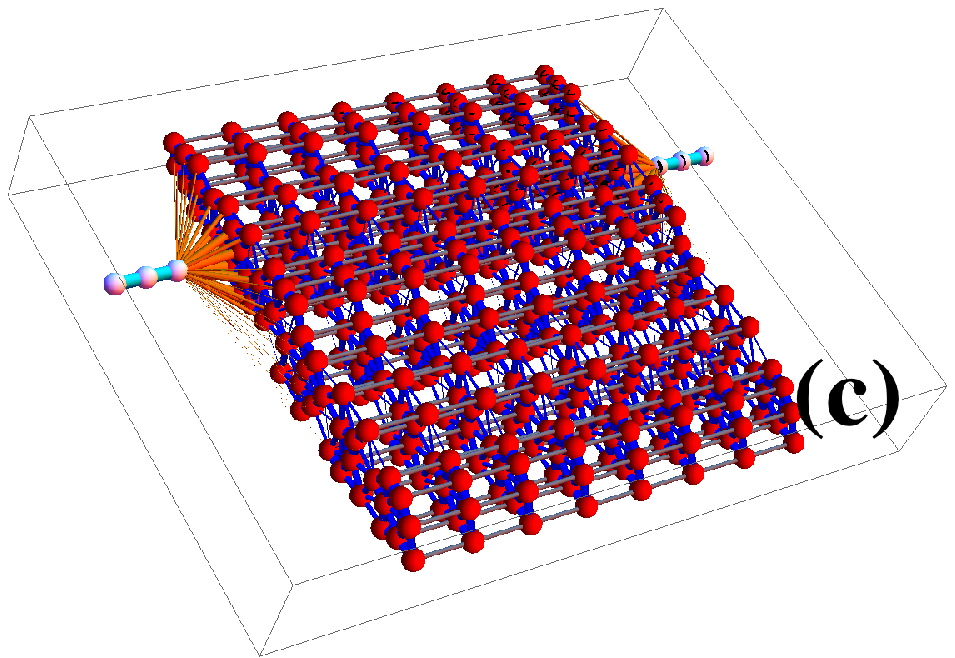}%   
\caption{\label{FigQManyDragon2} 
(Color online.)  
See the Appendix for full information on the figure construction. 
(a) A slice of a nanodevice made from random arrangements of 
atoms inside of an ellipse (green).  Here there are $m$$=$$50$ atoms in the slice.  
(b) The lead-device hopping terms $w_j$ required to make the nanodevice comprised of 
slices as in (a) into a quantum dragon.  The radii of the lead-slice 
(orange) cylinders are proportional to the strength of the required hopping terms.  
The positioning of the lead atom is at the CM of the $w_j$ bonds, but displaced 
above the slice along the direction of electron propagation in the leads.  
(c) The complete nanodevice formed from $\ell$$=$$8$ identical slices, connected 
to the lead atoms (white spheres) via the lead-slice hopping terms 
gives a quantum dragon.  
}
\end{figure}

This prescription does not require either that there be any 
symmetry within a slice or that the underlying graph for a slice is 
planar.  
Most importantly, we have not had to tune the connections or the 
intra-slice hopping strengths in a slice.  
The slice may be semi-regular, as in Fig.~\ref{FigQManyDragon1}, but that 
is not necessary.  One could have each slice, for example, be a portion of 
a two-dimensional quasi-crystal.  One can also have the slices atomic arrangement 
be completely random, as in an amorphous material.   
One example of such a complicated, amorphous nanodevice 
that can be a quantum dragon is shown in Fig.~\ref{FigQManyDragon2}.  
In this prescription, quantum dragons exist everywhere on a $2m+2$ dimensional 
\lq surface' in the $(m^2+5m+2)/2$ dimensional parameter space.   

%         %         %         %         %         %         %         %         %
%         %         %         %         %         %         %         %         %

%
%    %    %    %    %    %    %    %    %    %    %    %    %    %    %    %    %    %
%
\section{QUANTUM DRAGONS: PRESCRIPTION~2}

In the second prescription, we again assume that all slices are identical, 
and that only the simplest inter-slice hopping is present 
(${\bf B}$$=$$-s_0{\bf I}$).  
We further assume we have fixed arbitrary  
lead-device connections ${\vec w}_0$, and any 
fixed arbitrary intra-slice matrix ${\bf A}_0$.  
The subscript $0$ stands for the original given problem.  
We assume every element of ${\vec w}_0$ is negative, 
($-{\vec w}_0$$>$$0$).  
We are required to keep fixed the ratios of the elements of ${\vec w}_0$, 
but can tune the overall normalization.  Therefore the final lead-site hopping 
connections are given by 
\begin{equation}
{\vec w} = s_w {\vec w}_0
\end{equation}
where we can only tune $s_w$.  
Similarly, the connection to the output lead is given by 
\begin{equation}
{\vec u} = s_u {\vec w}_0
\end{equation}
where we can only tune $s_u$.  
Choose a coordinate system so the slice-to-slice direction of electron 
propagation is along the $z$-axis.  
We assume that the graph associated with ${\bf A}_0$ is strongly connected, 
even though ${\bf A}_0$ may have many elements equal to zero.  
We further assume we are only allowed to tune 
the intra-slice matrix by adding an electrical potential 
$V(x,y)$ at the location of every atom.  
The electric field is given in the standard fashion, 
${\vec E}(x,y)=-{\vec\nabla} V(x,y)$.  
The electric potential must be continuous, but otherwise its values at 
points between atoms does not enter the tight binding model.  
The electric field at the 
device location is the same for all slices.  If atom $j$ has the 
coordinates $(x_j,y_j)$, the $j^{\rm th}$ diagonal element of 
${\bf A}_0$ is thus changed by tuning the term $V(x_j,y_j)$.  
Introduce the $m$$\times$$m$ diagonal matrix ${\bf D}_V$ with 
the $j^{\rm th}$ diagonal element equal to the tuned electrical 
potential values, $V(x_j,y_j)$.  
The new intra-slice matrix with this added electrical potential is 
\begin{equation}
\label{Eq:AA0DV}
{\bf A} = {\bf A}_0 + {\bf D}_V
\>.
\end{equation}
Changing the electrical potential has not changed the intra-slice hopping 
strengths ($t_{j,j'}$) or the graph connectivity of a slice.  

We choose the electric potential so that $-{\bf A}\ge 0$, {\it i.e.\/} 
so all elements of ${\bf A}$ are non-positive.  
We want to choose a $\lambda_0$ and the $m$ diagonal elements of 
${\bf D}_V$ so that
\begin{equation}
\label{Eq:DvA0}
\left(-{\bf D}_V-{\bf A}_0\right){\vec w_0} = \lambda_0 {\vec w}_0
\end{equation}
and $-\left({\bf D}_V+{\bf A}_0\right)$$\ge$$0$.  
Introduce the vector ${\vec e}$ with all elements unity.  
Since the given ${\vec w}_0$ has no zero elements, there is a diagonal 
matrix ${\bf D}_e$ such that ${\bf D}_e{\vec w}_0={\vec e}$ and the 
inverse (diagonal) matrix ${\bf D}_e^{-1}$ exists.  
Also introduce the square matrix ${\bf J}$ with all elements unity.  
One has ${\bf J}$$=$${\vec e}{\vec e}^\dagger$, and for the ${\bf J}$ matrix 
$m$$\times$$m$ one has $\frac{1}{m}{\bf J}{\vec e}$$=$${\vec e}$.  
The diagonal matrix ${\bf D}_V$ we need to satiisfy Eq.~(\ref{Eq:DvA0}) is 
\begin{equation}
\label{Eq:DV}
{\bf D}_V = -\lambda_0{\bf I} \> - \>  
{\bf I} \odot \left({\bf D}_e{\bf A}_0{\bf D}_e^{-1}{\bf J}\right)
\end{equation}
where $\odot$ is the Hadamard (element-by-element) matrix product.  
The proof is that
\begin{equation}
\begin{array}{lcl}
{\bf D}_V {\vec w}_0 
& \> = \> & 
\left[
-\lambda_0{\bf I} - 
{\bf I} \odot \left({\bf D}_e{\bf A}_0{\bf D}_e^{-1}{\bf J}\right)
\right]{\vec w}_0
\\
& \> = \> & 
-\lambda_0{\vec w}_0 
- {\bf D}_e \left({\bf I}\odot\left({\bf A}_0 
{\bf D}_e^{-1}{\vec e}{\vec e}^\dagger\right)\right) {\vec w}_0 
\\
& \> = \> & 
-\lambda_0{\vec w}_0 
- {\bf D}_e \left({\bf I}\odot\left({\bf A}_0 
{\vec w}_0{\vec e}^\dagger\right)\right) {\vec w}_0 
\\
& \> = \> & 
-\lambda_0{\vec w}_0 
- \left({\bf I}\odot\left({\bf A}_0 
{\vec w}_0{\vec e}^\dagger\right)\right) {\bf D}_e {\vec w}_0 
\\
& \> = \> & 
-\lambda_0{\vec w}_0 
- \left({\bf I}\odot\left({\bf A}_0 
{\vec w}_0{\vec e}^\dagger\right)\right) {\vec e} 
\\
& \> = \> & 
-\lambda_0 {\vec w}_0 -{\bf A}_0{\vec w}_0
\\
\end{array}
\end{equation}
as required in order to satisfy Eq.~(\ref{Eq:DvA0}).  

Introduce the magnitude of 
${\vec w}_0$ to be $\left|{\vec w}_0\right|=\sqrt{{\vec w}^\dagger_0{\vec w}_0}$.  
Now choose the transformation matrix ${\bf X}$ to have the 
form as in Eq.~(\ref{Eq:Xw5}), with 
${\vec w}_1=-{\vec w}_0/\left|w_0\right|$, again with orthonormal vectors 
${\vec w}_i^\dagger{\vec w}_j=\delta_{j,j'}$ for $\{j,j'\}=1, 2, \cdots, m$.  
With this choice for the transformation matrix ${\bf X}$ 
the mapping equation Eq.~(\ref{Eq:Map3}) is satisfied with 
${\tilde s}_w=\left|w_0\right| s_w$.  Similarly, the mapping equation 
Eq.~(\ref{Eq:Map4}) is satisfied with 
${\tilde s}_u=\left|w_0\right| s_u$.  The lead-device connections will allow 
a quantum dragon if one chooses ${\tilde s}_w$$=$${\tilde s}_u$$=$$1$, 
which means we need to choose $s_w$$=$$s_u$$=$$1/\left|w_0\right|$.  

We have the simplest possible inter-slice hopping terms, having chosen 
${\bf B}=-s_0 {\bf I}$.  As in prescription~1, 
since ${\bf X}{\bf X}^\dagger$$=$${\bf I}$ the mapping equation Eq.~(\ref{Eq:Map2}) 
is satisfied.  The inter-slice terms allow a quantum dragon if $s_0$$=$$1$.  

With the tuned on-site energies  from ${\bf D}_V$ of 
Eq.~(\ref{Eq:DV}), the mapping equation 
Eq.~(\ref{Eq:Map1}) is satisfied with ${\tilde\epsilon}$$=$$-\lambda_0$.  
We are free to tune $\lambda_0$, since this would involve an equal shift of 
the electric potential on every site of the nanodevice.  
In order to have the intra-slice matrix 
allow a quantum dragon requires ${\tilde\epsilon}$$=0$, and therefore we tune to make 
$\lambda_0$$=$$0$.  From Eq.~(\ref{Eq:DV}) the required electric potential 
that must exist at each atomic site is 
\begin{equation}
\label{Eq:DVdragon}
{\bf D}_V = - \> 
{\bf I} \odot \left({\bf D}_e{\bf A}_0{\bf D}_e^{-1}{\bf J}\right)
\>.
\end{equation}

\begin{figure}[tb]
\includegraphics[width=5.1cm]{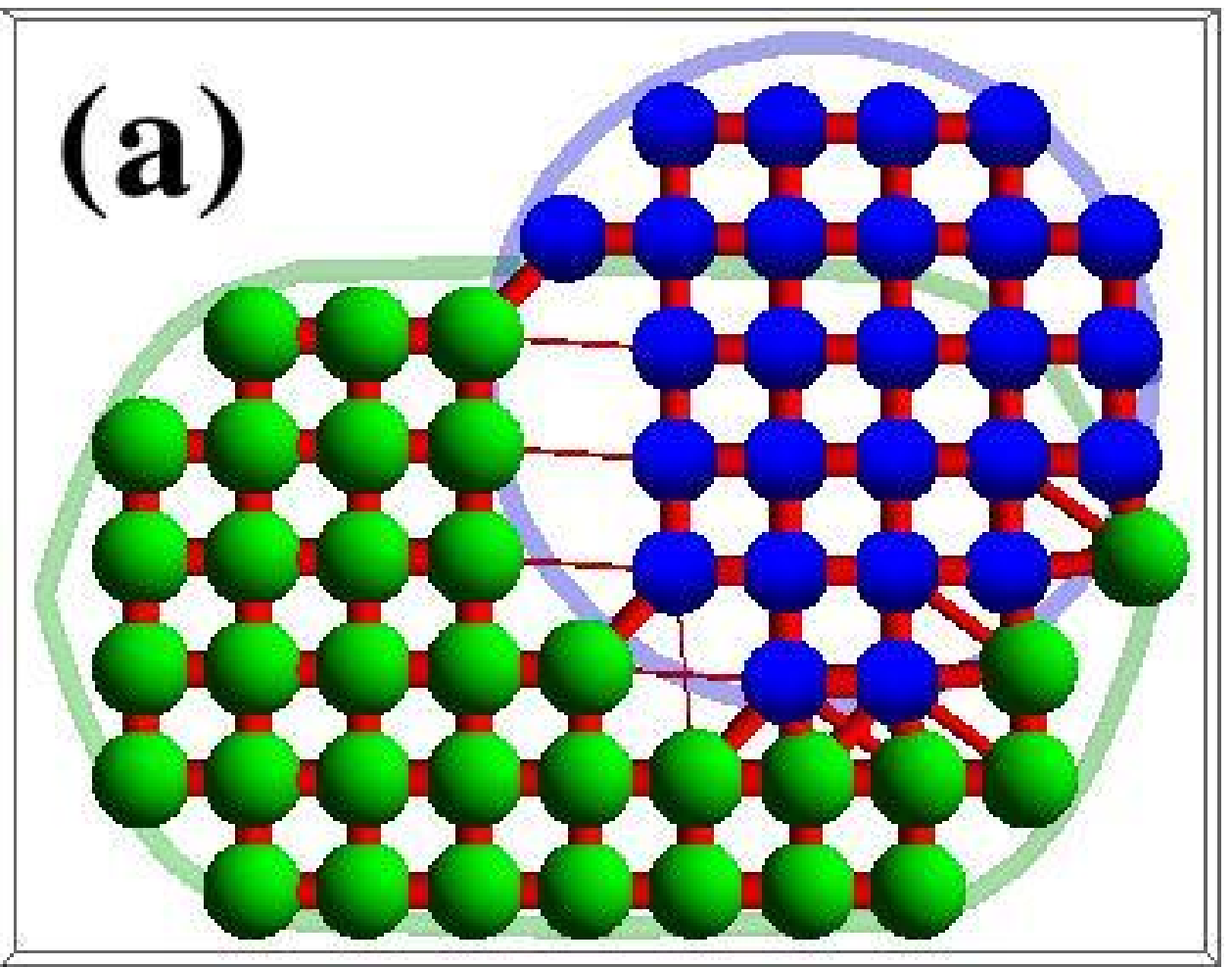}%   
\hskip 0.3 true in   
\includegraphics[width=5.1cm]{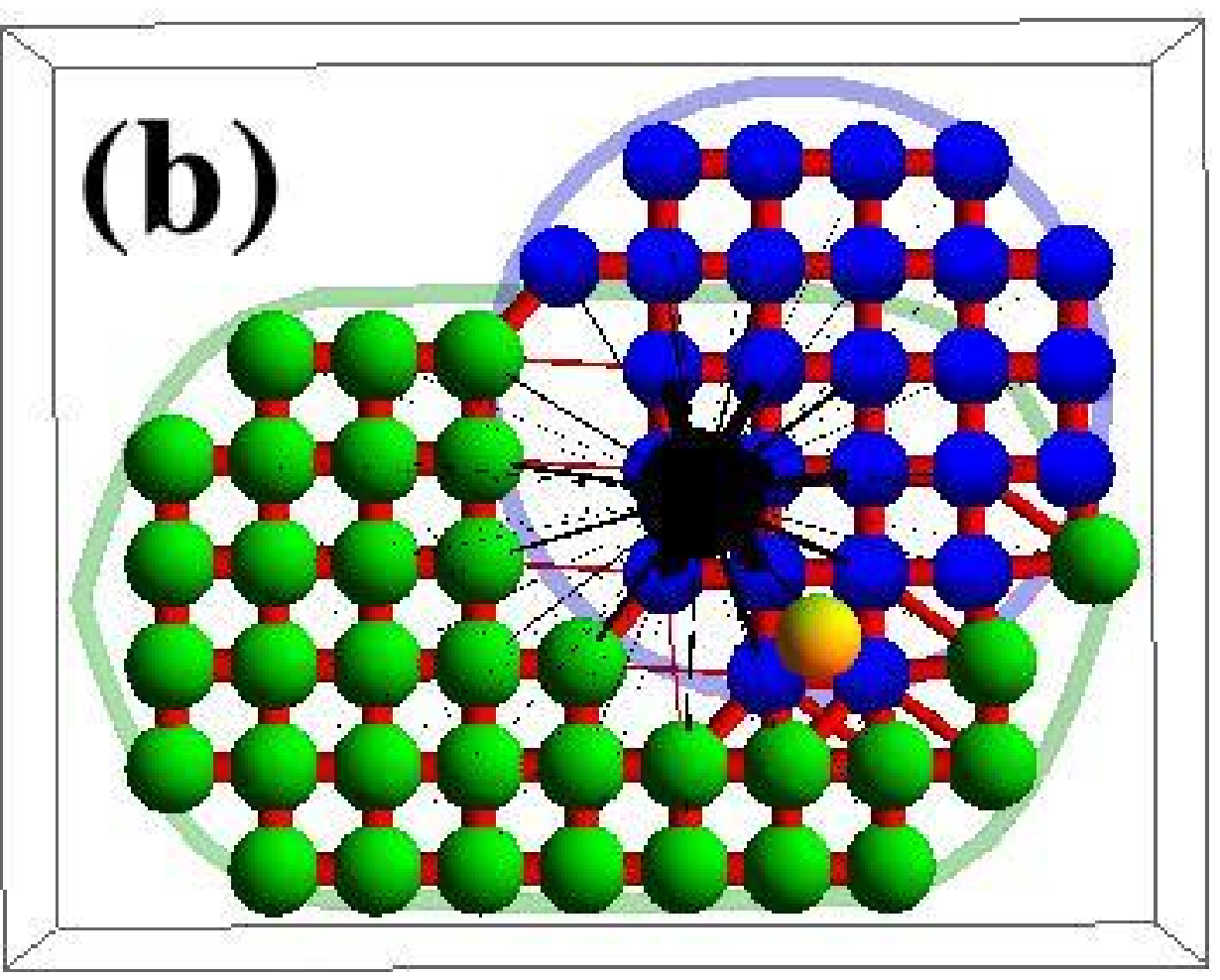}%   
\hskip 0.3 true in   
\includegraphics[width=5.1cm]{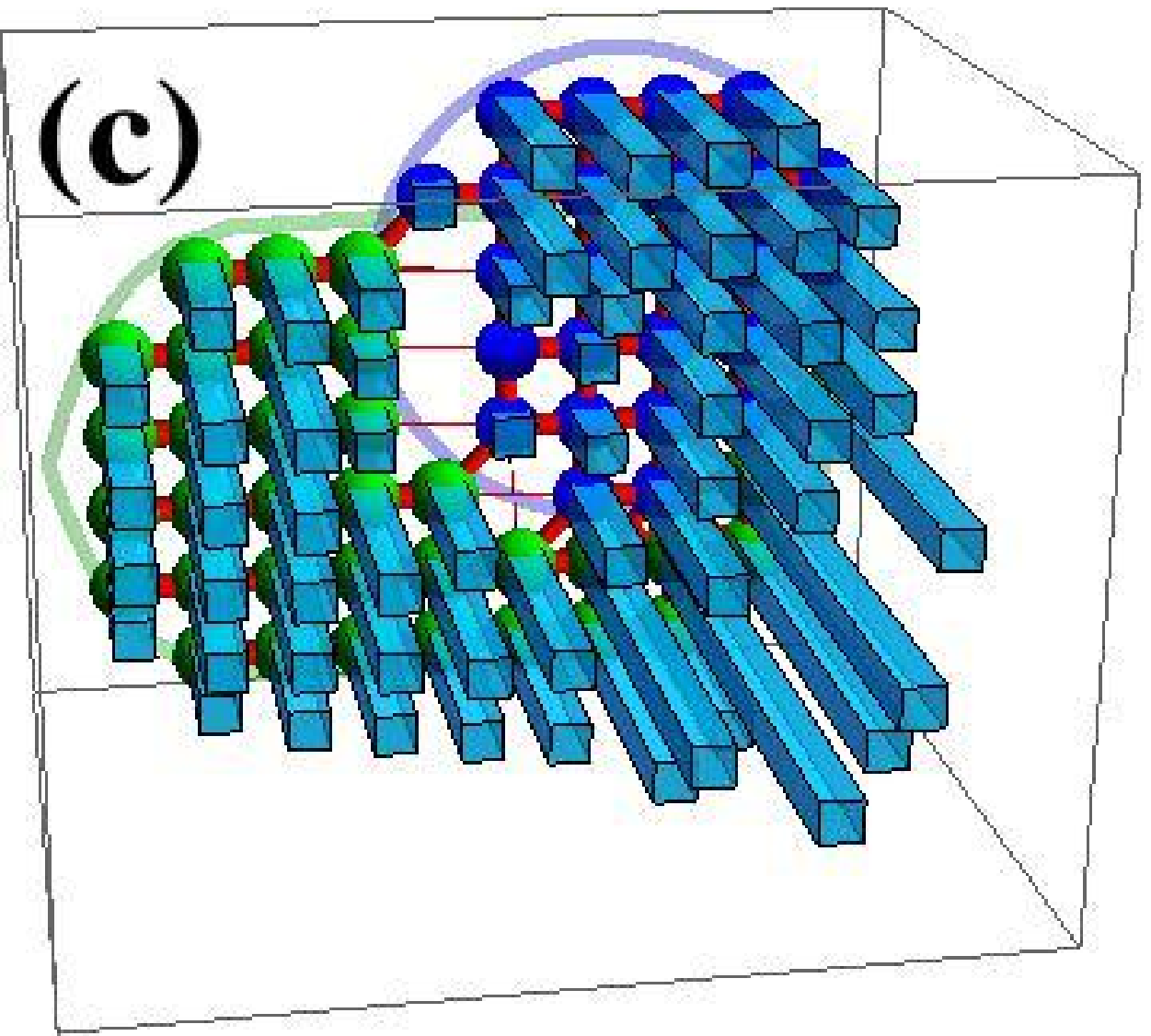}%   
\caption{\label{FigQManyDragon3} 
(Color online.)  
(a) A slice of a nanodevice made from square lattice arrangements of 
atoms inside of a generalized ellipse and a circle.  
There are $m$$=$$60$ atoms in the slice.  
(b) A lead-slice interaction connection, chosen without regard to the 
arrangement of atoms or bonds in the slice.  The (black) sphere 
is the last atom of the input lead and the (black) bonds are the hopping 
parameters (given by the vector ${\vec w}_0$) 
between this lead atom and the atoms in the first slice.  
The strength of the lead-slice interaction is proportional to the 
radius of the (black) cylinders.  
The location at which the lead atom would have to be placed in order to be at 
the CM of the ${\vec w}_0$ is shown as the yellow sphere.  
(c) The required electrical potential at each atomic site to make the 
device a quantum dragon.  The height of each (cyan) cuboid is proportional to 
the value of $V$ at the atomic site to make ${\cal T}(E)$$=$$1$.  For visual 
reasons, all potentials are shifted by a constant to make them all non-negative.  
}
\end{figure}

Therefore, we have found the solution to all four mapping equations in order 
to have a quantum dragon, {\it i.e.\/} to have ${\cal T}(E)$$=$$1$, for 
prescription~2.  We have tuned 
$2m$$+$$2$ tight binding parameters 
in order to have a quantum dragon, from the $(m^2+5m+2)/2$ total 
parameters [see Eq~(\ref{Eq:pres1Nterms})].  
An example of the tuning of each slice in prescription~2 is 
illustrated in Fig.~\ref{FigQManyDragon3}.  

In both prescriptions~1 and 2 the inter-slice mapping equation, Eq.~(\ref{Eq:Map2}), 
has been satisfied with the inter-slice matrix ${\bf B}=-s_0 {\bf I}$.  However, 
Eq.~(\ref{Eq:Map2}) could also be satisfied by having an inter-slice 
matrix of the form
\begin{equation}
\label{Eq:BIwwHomogeneous}
{\bf B}=-s_0 {\bf I} - 
s_{ww} \frac{{\vec w}_1{\vec w}_1^\dagger}{{\vec w}_1^\dagger{\vec w}_1}
\>.  
\end{equation}
In order to have a quantum dragon from the mapping, 
Eq.~(\ref{Eq:BIwwHomogeneous}) must be satisfied and we must tune 
the hopping strengths to satisfy $s_0$$+$$s_{ww}$$=$$1$.  
The third prescription is a generalization of this tuning of the 
inter-slice hopping terms to the case where every slice may be 
different.  

%         %         %         %         %         %         %         %         %
%         %         %         %         %         %         %         %         %

%
%    %    %    %    %    %    %    %    %    %    %    %    %    %    %    %    %    %
%
\section{QUANTUM DRAGONS: PRESCRIPTION~3; INHOMOGENEOUS SLICES}

We make similar assumptions about the intra-slice interactions as in 
prescription~1.  However, we now let every slice possibly be different.  
We assume the intra-slice matrices ${\bf A}_k$ for slices $1$$\le$$k$$\le$$\ell$ 
are given fixed values, 
and all slices are assumed to be strongly connected.  We assume we are again 
allowed to tune a constant electric potential $V_k$ by a 
constant value for every atom of slice~$k$.  
We assume we are allowed to tune the lead-device connection vector ${\vec w}$ 
for the input lead, and the lead-device connection vector ${\vec u}$ 
for the output lead.  We further assume that we are allowed to tune 
every element of the inter-slice hopping matrix 
${\bf B}_{k,k+1}$ for $k$$=$$1,2,\cdots (\ell-1)$.  

For this non-identical slice nanodevice, the matrix equation to analyze has the same 
form as Eq.~(\ref{Eq:EqN4}), with two differences.  The diagonal 
matrices (intra-slice terms) 
${\bf F}_k$$=$${\bf A}_k$$-$$E{\bf I}$ may all be different.  
The matrices ${\bf F}_k$ are $m_k$$\times$$m_k$, where slice $k$ 
has $m_k$ atoms (vertices).  
The inter-slice terms, ${\bf B}_{k,k+1}$ may all be different, and are in 
general not square matrices, being of size $m_k$$\times$$m_{k+1}$.  
In general, the device-output-lead interaction ${\vec u}$ will usually 
be tuned to be different from the device-input-lead interaction ${\vec w}$.  

For our non-identical slice case, to perform the mapping the transformation 
matrix ${\hat{\bf X}}$ of Eq.~(\ref{Eq:Xhat}) will have different unitary 
transformation matrices ${\bf X}_k$ of size $m_k$$\times$$m_k$ along the diagonal.  
The four mapping equations of Eq.~(\ref{Eq:Map1}) through Eq.~(\ref{Eq:Map4}) 
now become a set of $2\ell$$+$$1$ mapping equations.  The $\ell$ intra-slice 
mapping equations are 
\begin{equation}
\label{Eq:Map1non}
{\bf X}_k{\bf A}_k{\bf X}_k^\dagger = 
\left(
\begin{array}{cc}
{\tilde\epsilon}_k & {\vec 0}^\dagger \\
{\vec 0} & {\tilde {\bf A}}_k \\
\end{array}
\right)
\>.
\end{equation}
The $\ell$$-$$1$ inter-slice mapping equations are 
\begin{equation}
\label{Eq:Map2non}
{\bf X}_k{\bf B}_{k,k+1}{\bf X}_{k+1}^\dagger = 
\left(
\begin{array}{cc}
-{\tilde s}_{b,k} & {\vec 0}^\dagger \\
{\vec 0} & {\tilde {\bf B}}_k \\
\end{array}
\right)
\> .
\end{equation}
The lead-device hopping terms for the input lead must satisfy the mapping equation 
\begin{equation}
\label{Eq:Map3nonw}
{\bf X}_1{\vec w} = 
\left(
\begin{array}{c}
-{\tilde s}_w \\
{\vec 0} \\
\end{array}
\right)
\> ,
\end{equation}
while the lead-slice hopping terms for the output lead must satisfy the 
mapping equation
\begin{equation}
\label{Eq:Map3nonu}
{\bf X}_\ell{\vec u} = 
\left(
\begin{array}{c}
-{\tilde s}_u \\
{\vec 0} \\
\end{array}
\right)
\> .
\end{equation}
The $(m_k$$-$$1)$$\times$$(m_k$$-$$1)$ matrices ${\tilde{\bf A}}_k$ and 
the $(m_k$$-$$1)$$\times$$(m_{k+1}$$-$$1)$ matrices ${\tilde{\bf B}}_{k,k+1}$ 
are not important, since they will not be connected by any path to either the 
input or output leads \cite{QMdragon2014}.  

As for Eq.~(\ref{Eq:widetildeM4}), for the non-identical slice case we introduce 
the $(\ell$$+$$2)$$\times$$(\ell$$+$$2)$ matrix 
${\widetilde{\bf M}}_\ell$, written for $\ell$$=$$4$, as 
\begin{equation}
\label{Eq:widetildeM4non}
{\widetilde{\bf M}}_4 = 
\left(
\begin{array}{cccccc}
\xi(E) & -{\tilde s}_w & 0 & 0 & 0 & 0 \\
-{\tilde s}_w & {\tilde\kappa}_1 & -{\tilde s}_{b,1} & 0 & 0 & 0 \\
0 & -{\tilde s}_{b,1} & {\tilde\kappa}_2 & -{\tilde s}_{b,2} & 0 & 0 \\
0 & 0 & -{\tilde s}_{b,2} & {\tilde\kappa}_3 & -{\tilde s}_{b,3} & 0 \\
0 & 0 & 0 & -{\tilde s}_{b,3} & {\tilde\kappa}_4 & -{\tilde s}_u \\
0 & 0 & 0 & 0 & -{\tilde s}_u & \xi(E) \\
\end{array}
\right)
\end{equation}
with ${\tilde\kappa}_k={\tilde\epsilon}_k$$-$$E$.  
This is the matrix formed from 
only the $\ell$ transformed-sites 
after the $2\ell$$+$$1$ mapping equations 
Eq.~(\ref{Eq:Map1non}) through Eq.~(\ref{Eq:Map3nonu}) are used.  
The probability of transmission of the electron ${\cal T}=\left|t_T\right|^2$ 
is calculated from the $t_T$ quantities found from 
the solution of the matrix 
Eq.~(\ref{Eq:MappedMsmall}).  

To satisfy the intra-slice mapping equations 
in Eq.~(\ref{Eq:Map1non}), follow the same procedure based on the same 
arguments as in Eq.~(\ref{Eq:Rx1first}) to Eq.~(\ref{Eq:Rx1last}).  
In particular, introduce 
the maximum of zero or the 
largest positive diagonal element of ${\bf A}_k$, {\it i.e.\/} 
\begin{equation}
\epsilon_{{\rm max},k}={\rm max}\{0,\epsilon_{1,k},\epsilon_{2,k},\cdots,
\epsilon_{m_k,k}\}
\end{equation}
with the $\epsilon_{j,k}$ the on-site energy of the $j^{\rm th}$ 
atom in slice $k$.  
Introduce the matrix
\begin{equation}
{\hat{\bf A}}_k = \epsilon_{{\rm max},k} {\bf I} -{\bf A}_k
\> ,
\end{equation}
and hence ${\hat{\bf A}}_k\ge 0$.  
Let ${\vec w}_{1,k}$ be the normalized eigenvector of ${\hat{\bf A}}_k$ 
associated with 
the largest eigenvalue ${\hat\lambda}_{1,k}$ of ${\hat{\bf A}}_k$.
Since every ${\hat{\bf A}}_k$ has an associated strongly connected graph 
and is non-negative, 
by the extension of the Perron-Frobenius theorem 
\cite{MarcusMinc64,BermanPlemmons1979}
every vector ${\vec w}_{1,k}$ is unique and can be written with all 
non-negative values (${\vec w}_{1,k}$$>$$0$).  
Hence, ${\vec w}_{1,k}$ is also an eigenvector of ${\bf A}_k$ with eigenvalue 
$\lambda_{1,k}$.

We tune the lead-device hopping terms for the incoming lead, 
which must all be non-positive, to be 
\begin{equation}
{\vec w} = -{\hat s}_w {\vec w}_{1,1}
\end{equation} 
with some overall strength ${\hat s}_w$.
Similarly, we tune the lead-device hopping terms for the 
outgoing lead, which must be non-positive, to be 
\begin{equation}
{\vec u} = -{\hat s}_u {\vec w}_{1,\ell}
\end{equation} 
with some overall strength ${\hat s}_u$.
We have thus far had to form the matrices ${\hat{\bf A}}_1$ and 
${\hat{\bf A}}_\ell$ in order to use the Perron-Frobenius theorem to show the 
two lead-device vectors ${\vec w}$ and ${\vec u}$ are non-positive, as 
required for a physical tight-binding model.  

Choose the transformation matrices to be (written for $m_k$$=$$6$) 
\begin{equation}
\label{Eq:Xw6non}
{\bf X}_k^\dagger = 
\left(
\begin{array}{cccccc}
\> {\vec w}_{1,k} \>\> & 
{\vec w}_{2,k} \>\> & 
{\vec w}_{3,k} \>\> & 
{\vec w}_{4,k} \>\> & 
{\vec w}_{5,k} \>\> &
{\vec w}_{6,k} \>\> \\
\end{array}
\right)
\end{equation}
with orthonormal vectors ${\vec w}_{j,k}^\dagger {\vec w}_{j',k}$$=$$\delta_{j,j'}$.  
The vector ${\vec w}_{1,k}$ is an eigenvector of ${\bf A}_k$.  The 
other $m_k$$-$$1$ vectors ${\vec w}_{j,k}$ for $j$$=$$2, 3, \cdots, m_k$ 
need not be eigenvectors of the ${\bf A}_k$, only orthonormal to each other and 
to ${\vec w}_{1,k}$.  
With this choice of ${\bf X}_k$ and the tuned values for the 
lead-device vectors ${\vec w}$ and ${\vec u}$, 
the mapping equations (\ref{Eq:Map3nonw}) and (\ref{Eq:Map3nonu}) are 
satisfied with ${\tilde s}_w$$=$${\hat s}_w$ 
and ${\tilde s}_u$$=$${\hat s}_w$.  

We apply a constant electrical potential $-V_{{\rm shift},k}$ 
to every atom in the slice $k$ of the nanodevice, so every matrix 
${\bf A}_k$ is shifted to the matrix ${\bf A}_k$$-$$V_{{\rm shift},k}{\bf I}$.  Then 
\begin{equation}
\left(
{\bf A}_k - V_{{\rm shift},k}{\bf I}
\right) {\vec w}_{1,k} =
\left(
\lambda_{1,k} - V_{{\rm shift},k}
\right) {\vec w}_{1,k} 
\> .
\end{equation}
With our choice of ${\bf X}_k$, the mapping equation Eq.~(\ref{Eq:Map1non}) is 
satisfied with 
\begin{equation}
{\tilde\epsilon}_k = 
\lambda_{1,k}(\epsilon_{{\rm max},k}) - V_{{\rm shift},k}
\>.
\end{equation}

In order to satisfy the mapping equations in Eq.~(\ref{Eq:Map2non}) we tune 
the inter-slice hopping matrices, which are $m_k$$\times$$m_{k+1}$, to be
\begin{equation}
{\bf B}_{k,k+1} 
\> = \> 
- s_{b,k} {\vec w}_{1,k} {\vec w}_{1,k+1}^\dagger
\end{equation}
with arbitrary inter-slice hopping strengths $s_{b,k}$.  
We have had to form the matrices ${\hat{\bf A}}_k$ 
in order to use the Perron-Frobenius theorem to show that all 
elements of ${\bf B}_{k,k+1}$ are non-positive.  

All $2\ell$$+$$1$ mapping equations for the non-identical slice case 
have thus been satisfied.  
For a quantum dragon using this prescription, we need only further tune the 
values so 
\begin{equation}
\begin{array}{lcl}
{\vec w} & \> = \> & - {\vec w}_{1,1} \> , \\
{\vec u} & \> = \> & - {\vec w}_{1,\ell} \\
V_{{\rm shift},k} & \> = \> & \lambda_{1,k} \\
& {\rm and} & \\
{\bf B}_{k,k+1} & \> = \> & - {\vec w}_{1,k} {\vec w}_{1,k+1}^\dagger \\
\end{array}
\end{equation}
so that ${\tilde s}_w$$=$$1$, ${\tilde s}_u$$=$$1$, 
${\tilde s}_{b,k}$$=$$1$, and ${\tilde\epsilon}_k$$=$$0$ for all 
slices $1$$\le$$k$$\le$$\ell$.  

\begin{figure}[tb]
\includegraphics[width=7.3cm]{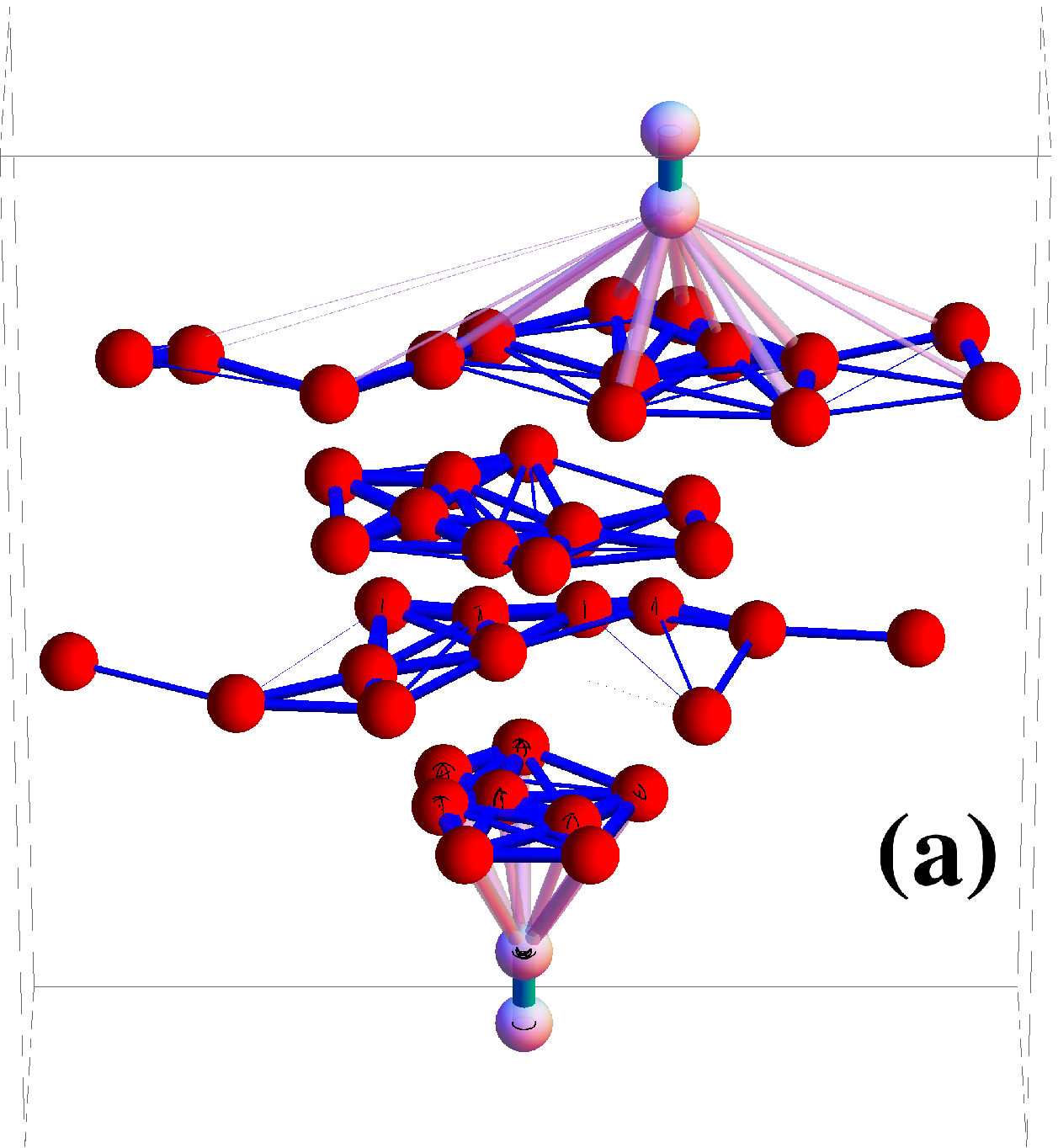}%   
\hskip 1.3 true in   
\includegraphics[width=7.3cm]{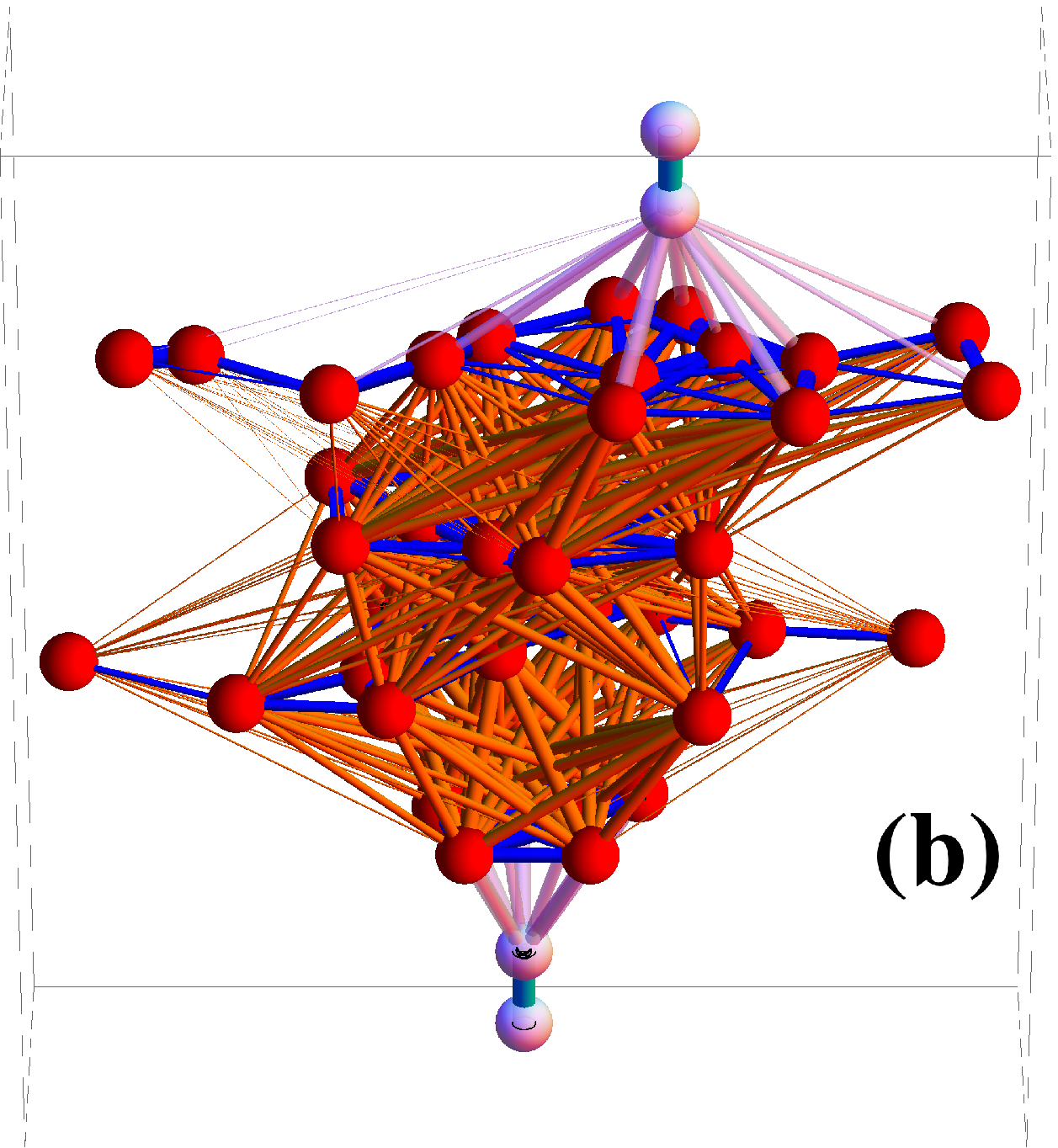}%   
\caption{\label{FigQManyDragon4} 
(Color online.)  
A nanodevice which can be a quantum dragon.  The device 
has a total of $44$~atoms in $\ell$$=$$4$ slices.  
The widths of the cylinders are proportional to the size of the hopping parameters. 
See the appendix for a full description.  
(a) Showing only the device atoms, lead atoms (two in each lead) and bonds, 
lead-device interactions, 
and intra-slice interactions.  
The lead atoms are positioned at the CM locations given by $w_j$ and $u_j$ for the 
first and last slice, respectively.  
(b) The figure in (a), with the inter-slice tuned hopping interactions 
(orange cylinders) 
in prescription~3 to make the device a quantum dragon.  
}
\end{figure}

Each intra-slice matrix ${\bf A}_k$ is symmetric, and hence has 
$m_k(m_k+1)/2$ tight-binding parameters.  We have had to tune the 
diagonal $m_k$ of these 
by the same amount $V_{{\rm shift},k}$, so each slice has the one 
parameter $V_{{\rm shift},k}$ that needs to be tuned.  There is only a unique 
choice for all inter-slice hopping terms ${\bf B}_{k,k+1}$ and the 
two lead-slice hopping terms ${\vec w}$ and ${\vec u}$.  
The free tight-binding parameters are only from the 
intra-slice matrices ${\bf A}_k$, for a total number of arbitrary parameters 
\begin{equation}
\label{Eq:Pres3Ntune}
\begin{array}{lcl}
N_{\rm free} & \>=\> & 
\sum_{k=1}^\ell \left[\frac{m_k\left(m_k+1\right)}{2}-1\right] \\
& \>=\> & \sum_{k=1}^\ell \frac{\left(m_k+2\right)\left(m_k-1\right)}{2}
\> . 
\end{array}
\end{equation}
The total number of tight-binding parameters is  
\begin{equation}
\label{Eq:Pres3Ntot}
N_{\rm Tot} = 
m_1 
\> + \> 
\sum_{k=1}^\ell \frac{m_k\left(m_k+1\right)}{2}
\> + \> 
\sum_{k=1}^{\ell-1} m_k m_{k+1}
\> + \> 
m_\ell
\> .
\end{equation}
An example of the formation of a quantum dragon using 
prescription~3 is shown in Fig.~\ref{FigQManyDragon4}.  
Prescription~3 does not require any translational symmetry, either 
along the direction of electron motion in the leads or perpendicular to 
this axis.  Therefore, the nanodevice can be considered to be 
completely disordered.  Nevertheless, all incoming electrons will be 
completely transmitted through the nanodevice, {\it i.e.\/} it is a 
quantum dragon having ${\cal T}(E)$$=$$1$.  The quantum dragons exist 
only on a low-dimensional \lq surface' [dimension given by Eq.~(\ref{Eq:Pres3Ntune})]
in the entire tight-binding 
parameter space [dimension given by Eq.~(\ref{Eq:Pres3Ntot})].  

%         %         %         %         %         %         %         %         %
%         %         %         %         %         %         %         %         %

%
%    %    %    %    %    %    %    %    %    %    %    %    %    %    %    %    %    %
%
\section{DISCUSSION AND CONCLUSIONS}

We have shown that quantum dragons are ubiquitous.  They exist for {\it any\/} 
fixed atomic bonding arrangement!  
We have chosen to concentrate on nanodevices with 
$\ell$ slices, but the prescriptions also work when there is only $\ell$$=$$1$ slice. 
For $\ell$$=$$1$, any atomic bonding arrangement between the atoms is possible.  
The only question in all prescriptions is how many 
tight-binding parameters need to be tuned, and to what values these parameters 
must be tuned.  
The prescribed tight-binding parameters must be tuned in order to 
satisfy the mapping equations Eq.~(\ref{Eq:Map1}) through Eq.~(\ref{Eq:Map4}).  
To allow the nanodevice to be a quantum dragon requires further tuning to 
specific tight-binding values.  With careful tuning,  electrons of all energies will 
have complete transmission, ${\cal T}(E)$$=$$1$, and the device will hence be 
a quantum dragon.  

Three prescriptions, all allowing for quantum dragons from 
inhomogeneous nanodevices, have been presented in detail.  
The first two prescriptions have the simplest inter-slice hopping terms, 
and have every slice identical.  
The first prescription allows arbitrary fixed values for the positions of all atoms, 
of all electrical potentials (up to a constant shift), 
and of all the intra-slice hopping strengths between atoms.  
In the first prescription the lead-device hopping strengths must be tuned in a 
prescribed manner in order to obtain a quantum dragon.  
The second prescription is related to the first, but the lead-device connections 
are arbitrary values (but identical connections for the input lead and output lead), 
while the electrical potentials on each atom must be tuned in a prescribed manner 
in order to obtain a quantum dragon.  
The third prescription has slices which may all be different, with atomic 
bonding strengths and electrical potentials (up to a slice-dependent constant term) 
fixed arbitrarily, while 
the lead-device and inter-slice hopping terms must be tuned in a prescribed 
manner in order to obtain a quantum dragon.  

For all prescriptions the number of arbitrary parameters is much larger than 
the number of parameters which must be tuned in a particular fashion.
For the first two prescription, the ratio of the number of tight-binding parameters 
is 
\begin{equation}
\frac{\rm tuned}{\rm Total} = 
\frac{4\left(m+1\right)}{m^2+5m+2}
\end{equation}
where $m$ is the number of atoms in every slice of the nanodevice.  
For the third prescription the number of atoms $m_k$ in each of the 
$1$$\le$$k$$\le$$\ell$ slices 
can be different.  However, if all $m_k$$=$$m$ (but the intra-slice bonds and 
electrical potential may be different for every slice) the ratio of the number of 
tight-binding parameters is
\begin{equation}
\frac{\rm tuned}{\rm Total} = 
\frac{\ell m^2 + \ell m -2\ell}{3\ell m^2+\ell m -2m^2+4m}
\>.  
\end{equation}
Thus in all cases, 
quantum dragons exist only on a low-dimensional \lq surface\rq ~ of the 
high-dimensional tight-binding parameter space.  
An analogy might be useful to understand the relationship between the 
complete tight-binding parameter space, the parameter space of the mapping equations, 
and the parameter space of quantum dragons.  Consider a room, so the space has 
dimension $D$$=$$3$, which can be viewed as the complete parameter space 
for this analogy.  A thin sheet of paper in the room, maybe folded or crumpled, 
represents the parameter space where the mapping equations hold, here $D$$=$$2$.  
A curve drawn on the sheet of paper represents the parameter space where 
quantum dragons exist, here $D$$=$$1$. 
Clearly, a blind Monte Carlo search of the $D$$=$$3$ space would have 
zero probability of locating a point exactly on the $D$$=$$2$ surface, 
much less on the $D$$=$$1$ curve.  

The natural question is which, if any, prescription would yield a 
nanodevice and lead-slice connections that can be realized reasonably easily 
experimentally.  The answer is that all three prescriptions have 
experimental difficulties.  The first prescription can be used to have a 
single-slice ($\ell$$=$$1$) nanodevice, and a quantum dragon can always be found 
with the correct lead-device connections.  However, to make the lead-slice 
connections would be physically impossible at the nanoscale, particularly 
for a non-planar arrangement of atoms in the slice.  The second prescription 
for completely arbitrary lead-device connections would require electric 
fields on the order of $10^9$~V/m precisely tuned at the nanoscale level.  
Although not physically impossible, such high electric fields tuned to 
the nanoscale level would set an extremely high experimental bar.  
The third prescription requires inter-slice connections that seem impossible, 
even for the case of Fig.~\ref{FigQManyDragon4} with only about ten atoms per slice.  

A more realistic method of experimentally constructing an inhomogeneous 
quantum dragon might take an approach that is a combination of the three 
prescriptions.  For example, one could require that the lead-device interactions 
should be monotonically dependent on the distance from the lead atom, and the 
electric field required to change by no more than a few percent on 
the nanoscale level.  These types of smoothness constraints are not necessary 
mathematically, but will be critical to synthesizing an experimental realization 
of inhomogeneous quantum dragons.  With such smoothness constraints, a 
physical nanodevice may be synthesized experimentally using a combination of 
the first two prescriptions.  

The structure of the nanodevice in Fig.~\ref{FigQManyDragon1} in particular 
seems like this type of quantum dragon should be amenable to experimental 
synthesis.  The metal polonium (Po) has a simple cubic 
lattice structure, and hence Fig.~\ref{FigQManyDragon1} can be viewed as a 
nano-crystal of Po.  Furthermore, the proofs show that any nano-crystal of 
Po can be connected to be a quantum dragon, particularly with the 
end slices as (100) faces as in Fig.~\ref{FigQManyDragon1}.  Of course, given the 
half-life of ${}^{208}$Po is about 2.9~y and that of ${}^{209}$Po is about 
125~y, a perfect nano-crystal of Po will only survive so long before nuclear 
decay creates defects.  It should be possible to connect homogeneous leads 
to a Po nano-crystal, and the search for electrical conductivity 
showing dragon segments may be helped by shaped electric potentials in the 
nanocrystal.  

The proofs of existence of quantum dragon segements here 
are only for leads of a single channel, 
and for homogeneous leads.  
Furthermore, the proofs are for the single-band tight binding model.  
It may be possible to extend these proofs to 
multi-channel leads, to other more complicated leads, and to more realistic 
band models.  Of particular interest 
would be to try to find quantum dragons in face-centered cubic single crystal 
nanodevices.  This paper has shown how ubiquitous quantum dragons are for 
the simplest cases, but lends hope to their existence in more complicated 
nanodevices connected to more complicated leads.  

It is anticipated that quantum dragons will at least have similar technological 
applications as do ballistic electron propagation 
devices \cite{Hanson2008,Javey2003,Wu2012,Kim2014}.  
Previously all known electron propagation with complete electron 
transmission, ${\cal T}(E)$$=$$1$, were for homogeneous nanodevices.  
Whether quantum dragons that are inhomogeneous will enable additional 
technological applications is an active topic of research.  

%\vfill\eject

\noindent{\bf Acknowledgements}  
Useful conversations are acknowledged with 
O.\ Abdurazakov, 
H.\ De.~Raedt, 
G.\ Inkoom, 
F.\ Jin, 
Z.\ Li, 
K.\ Michielsen, 
T.\ Neuhaus, 
P.A.\ Rikvold, 
and 
L.\ Solomon.  
Supported in part by US National Science Foundation grant DMR-1206233.  
Hospitality of the J{\"u}lich Supercomputing Centre (JSC) in J{\"u}lich, Germany 
is gratefully acknowledged.  

%\vfill\eject
%%%%%%%%%%%%%%%%%%%%%%%%%%%%%%%%%%%%%%%%%%%%%%%%%%%%%%%%%%%%%%%%%%%%%%%%%%%%%%%%%%%%%%

%
%    %    %    %    %    %    %    %    %    %    %    %    %    %    %    %    %    %
%
\appendix
\section*{Appendix}

Herein details describing the four figures are given. Although the figures are 
somewhat schematic, being only four examples of quantum dragons, it is 
informative nevertheless to give details of their construction.  
The length units are relative since the figures are schematic, 
but due to the atomic nature of the nanodevice, 
are expected to be about a nanometer.  
All figures were made in Mathematica \cite{Mathematica}.

{\bf Figure~1.}~ 
Fig.~\ref{FigQManyDragon1} shows the construction of 
a quantum dragon nanodevice using prescription~1 for a 
nano-crystal cut from a simple cubic lattice.  
The slice in Fig.~1(a) is made by keeping only sites from a 
square lattice within an ellipse (green).  
The square lattice has lattice spacing unity, and each atom is a (red) 
sphere of radius~0.5.  Both nearest-neighbor 
(which cannot be seen because the spheres touch) and next-nearest hopping 
terms (blue cylinders) are present.  Only the $m$$=$$76$ atoms within 
an ellipse centered at the randomly chosen point $(0.539,0.253)$ with 
axis along $x$ of length~12 and along $y$ of length~2 are used in the slice.
The number of non-zero hopping terms (number of bonds) in the slice is~236.  
Fig.~1(b) shows the connections between the lead (white sphere) and the first 
slice of the nanodevice, with the slice being identical to that of Fig.~1(a).  
The radius of the lead-device (orange) cylinders are proportional to the 
strength of a particular lead-device hopping strength $s_w$.  
The lead-device hopping values are found by finding a vector 
${\vec w}_1$ which is an eigenvector of ${\bf A}$ with all non-positive 
elements, which is guaranteed to exist by the Perron-Frobenius theorem.  
The input lead atom is placed at the CM given by the $w_j$, but positioned above the 
first slice.  
Fig.~1(c) shows the completed nanodevice.  Only two atoms (white spheres) 
in each semi-infinite lead are shown, and for clarity the lead bonds 
(cyan cylinders) are plotted at a distance of~1.5.  
The end atom of each semi-infinite lead is plotted to be separated from the 
blob by~1.5.  
There are $\ell$$=$$20$ slices plotted, separated by a distance of~1.1.  
In Fig.~\ref{FigQManyDragon1}, since $m$$=$$76$, 
quantum dragon segments exist on a 154-dimensional \lq surface' 
in 3079-dimensional space of all tight-binding parameters.  

{\bf Figure~2.}~ Fig.~\ref{FigQManyDragon2} shows the construction of 
a quantum dragon nanodevice using prescription~1 for a randomly constructed slice.  
(a)~The arrangement of the atoms (red spheres) in the slice is constructed by 
choosing an ellipse (green curve) 
centered at the origin with an $x$-axis equal to 12 and a 
$y$-axis equal to 2.  Fifty atoms are placed with uniform probability within 
the ellipse.  The atoms are assumed to have a hard radius equal to 0.5, 
and hence cannot be placed closer than a center-to-center distance of 
$d_{j,j'}$$=$$1.0$.  
The intra-slice bonds are shown by (blue) cylinders, with bonds placed between 
any atoms with $d_{j,j'}$$\le$$2.5$.  The strength 
of the intra-slice bonds are given by a linear relationship in the 
center-to-center distance, with bond strength unity for $d_{j,j'}$$=$$1$ and 
zero for $d_{j,j'}$$=$$2.5$.  
(b)~The connection of a slice to a lead atom (white sphere) in order to 
satisfy prescription~1 is shown.  
The lead-device hopping strengths are calculated by finding a vector 
${\vec w}_1$ which is an eigenvector of the intra-slice ${\bf A}$ 
with all non-positive elements.  The vector ${\vec w}_1$ 
is unique and guaranteed to exist by the Perron-Frobenius theorem.  
The strength of the lead-slice bonds are proportional to 
the radii of the (orange) cylinders.  
The input lead atom is positioned at the CM given by the hopping 
parameters $w_j$, but above the plane of the first slice.  
(c)~The complete nanodevice, here composed of $\ell$$=$$8$ identical slices.
The transmission can be calculated either from 
Eq.~(\ref{Eq:MappedMbig}) which requires finding the inverse of a 
$(\ell m$$+$$2)$$\times$$(\ell m$$+$$2)$$=$$402$$\times$$402$ matrix or from 
Eq.~(\ref{Eq:MappedMsmall}) which requires finding the inverse of a
$(\ell$$+$$2)$$\times$$(\ell$$+$$2)$$=$$10$$\times$$10$ matrix.  
Fig.~\ref{FigQManyDragon2} has $m$$=$$50$, 
dragon segments exist on a 102-dimensional \lq surface' 
in the 1376-dimensional space of all tight-binding parameters.  

{\bf Figure~3.}~ Fig.~\ref{FigQManyDragon3} shows the construction of 
a slice of a quantum dragon nanodevice using prescription~2.  
The lattice spacing is set to one for both the green and blue square lattices.  
(a) The (green sphere) atoms are within 
a generalized ellipse with the equation 
\begin{equation}
\left(\frac{x}{5}\right)^6 + \left | \frac{y}{3}\right| \> \le \> 1
\end{equation}
with the center randomly set.  All 34 (green) atoms are inside this (green) 
generalized ellipse but not inside the blue circle of radius~3 centered at the 
point $(2,2)$.  The 26 (blue) atoms are inside this (blue) circle.  The square 
lattice of these (blue) atoms is randomly offset from the center.  Any of the 
(blue) atoms inside the circle that would be at a distance less than unity to 
a (green) atom are not included.  The slice has $m$$=$$60$ atoms.  
Green-to-green atom bonds (50 bonds) and blue-to-blue 
atom bonds (40 bonds) are only between atoms at a 
distance of unity (nearest neighbors).  
Green-to-blue atoms bonds are inversely proportional to their length, and 
are between the atoms at a distance of less than two, giving 26 such bonds.  
Therefore there is a total of 116 intra-slice bonds (orange cylinders).  
(b) The yellow sphere shows the position of the lead atom in order for it to 
be at the center of mass [located at (2.08, -0.33)] 
of the slice for the eigenvector of ${\bf A}$ (as in prescription~1).  
Instead, the lead atom is chosen to be at the point (1,1,2), given by the 
(black) sphere, while the slice is in the plane $z$$=$$0$.  
Every atom in the slice is connected to 
the lead atom with a strength (black cylinders) chosen to be proportional to 
$r_j^{-6}$, with $r_j$ the distance between the lead atom and the $j^{\rm th}$ 
atom in the slice.  
(c) The required electric potential $V_j$ at every atom site $j$ in the 
slice is shown (cyan cuboid) 
in order to make the slice satisfy the mapping of Eq.~(\ref{Eq:Map1}). 
For visual reasons, these potentials are all shifted by the same amount in 
order to make them all non-negative.  For a quantum dragon, typically the 
required $V_j$ may be of different signs for different atoms.  
In Fig.~\ref{FigQManyDragon3}, since $m$$=$$60$, 
dragons live on a 122-dimensional \lq surface' 
in the tight-binding 1951-dimensional space.  

{\bf Figure~4.}~ 
Fig.~\ref{FigQManyDragon4} shows the construction of 
a quantum dragon nanodevice using prescription~3 for completely random 
inhomogeneous slices.  
The figure has $\ell$$=$$4$ slices.  The ellipse for the random 
placement of atoms in each slice has the $x_{\rm axis}$ in Table~1 and 
$y_{\rm axis}$$=$$2$.  The spheres have a hard core radius~$\frac{1}{2}$, and 
the intra-slice bonds are for any atom pairs in the same slice with 
a distance less than~$4$.  
Table~1 shows the specific values for each slice.
The fixed intra-slice bond strengths are proportional to the width of 
the intra-slice (blue) cylinders, and were chosen as a linear function 
of distance, with width one for $d_{j,j'}$$=$$1$ and width zero 
for $d_{j,j'}$$=$$4$.  
The diameters of the cylinders of the inter-slice bonds, 
and of the lead-device bonds, are proportional to the tuned bond strengths 
required for the device to be a quantum dragon.  
The input (output) lead atoms are placed at the CM of the 
hopping parameters $w_{1,1}$ ($w_{1,\ell}$) that connect the input (output) 
lead to the atoms in the first (last or $\ell^{\rm th}$) slice, but 
just below (above) the first (last) slice.  
In Fig.~\ref{FigQManyDragon4}, since $\ell$$=$$4$ and the $m_k$ are listed 
in Table~1, 
the quantum dragons exist on a 270-dimensional \lq surface' 
in the 652-dimensional space of all tight-binding parameters.  

\begin{table}[tbh]
\caption{Parameters related to Fig.~\protect\ref{FigQManyDragon4}}
\begin{ruledtabular}
\begin{tabular}{|c|r|r|c|}
Slice \# & $m_k$ & $x_{\rm axis}$ & \# intra-slice \\
 $k$ & & & bonds \\
\hline
1 &  8 &  2.0 & 28 \\
2 & 12 &  8.0 & 27 \\
3 & 10 &  4.0 & 33 \\
4 & 14 & 10.0 & 38 \\
\end{tabular}
\end{ruledtabular}
\end{table}

%%%%%%%%%%%%%%%%%%%%%%%%%%%%%%%%%%%%%%%%%%%%%%%%%%%%%%%%%%%%%%%%%%%%%%%%%%%%%%%%

\end{document}